\documentclass[aps,showpacs]{revtex4}
\usepackage[colorlinks=true, pdfstartview=FitV, linkcolor=blue, citecolor=red, urlcolor=magenta, breaklinks=true]{hyperref}
\usepackage{color}
\usepackage{subeqnarray}
\usepackage{graphicx}
\usepackage[all]{xy}
\usepackage{amsmath}
\usepackage{amssymb}
\newcommand{\be}{\begin{equation}}
\newcommand{\ee}{\end{equation}}
\newcommand{\ben}{\begin{eqnarray}}
\newcommand{\een}{\end{eqnarray}}
\newcommand{\bes}{\begin{subequations}}
\newcommand{\ees}{\end{subequations}}
\newcommand{\bens}{\begin{subeqnarray}}
\newcommand{\eens}{\end{subeqnarray}}
\newcommand{\wt}{\widetilde}
\newcommand{\rt}{\right}
\newcommand{\lt}{\left}
\newcommand{\bb}{\bibitem}
\def\tanh{\text{tanh}}
\def\arcsinh{\text{arcsinh}}
\def\sech{\text{sech}}
\def\min{\text{min}}
\def\max{\text{max}}
\begin{document}
\title{Novel results for kinklike structures and their connections to quantum mechanics}
\author{D. Bazeia$^1$}
\author{D.A. Ferreira$^1$}
\author{Elisama E.M. Lima$^{1,2}$}
\author{L. Losano$^1$}
\affiliation{{\small $^1$Departamento de F\'{\i}sica, Universidade Federal da Para\'{\i}ba\\
Caixa Postal 5008, 58051-970 Jo\~ao Pessoa, Para\'{\i}ba, Brazil}\\{$^2$Instituto Federal do Rio Grande do Norte, 59500-000 Macau, Rio Grande do Norte, Brazil}}
\vspace{3cm}
\date{\today}

\begin{abstract}

In this work we use the deformation procedure and explore the route to obtain distinct field theory models that present similar stability potentials. Starting from systems that interact polynomially or hyperbolically, we use a deformation function which allows to construct other theories having the feature of giving rise to the same stability potentials. Such deformation function leads to smooth potentials according to a specific choice of a single parameter. Among the results, one shows that for models with asymmetric topological sectors, the appearance of a new stability potential is also possible.

\end{abstract}


\maketitle

\section{Introduction}

Topological defects are localized structures that appear in a diversity of contexts \cite{RA,RS,VS,MS,V} in high energy physics.
Among the most popular, stand out the one-dimensional structures called kinks which may provide important informations about the behavior of several physical systems; see, e.g., Refs.~\cite{RA,RS,VS,MS,V,Mussardo,Basar,Izquierdo, Romanczukiewicz,Pyka,Correa,Dupuis}. In the case of field theories, such structures appear generally  as solutions of nonlinear equations of motion and present localized energy density and finite energy. 

An interesting fact about kinks concerns their linear stability, whose analysis leads to a direct connection with supersymmetry and quantum mechanics \cite{qm1,qm2}. This connection is an old issue and has been investigated in several works, in particular in Refs.~\cite{OI1,OI2,OI3,OI4,OI5,BO,VA,BEN1,BEN2,BL}. Along the years, it has been shown that the subject engenders two interesting routes, one that goes from field theory to quantum mechanics, when one investigates stability of the kinklike structures in a field theory model, which leads us to a stability potential that simulates a quantum mechanical potential. The other route is referred to as the reconstruction procedure, in which a field theory model is constructed from specific features of a given quantum mechanical potential, in particular its zero or translational mode. In the works \cite{BO,VA} the authors presented interesting investigations concerning the reconstruction of field theories from the spectrum of quantum mechanical potentials, and these studies have motivated the recent Refs.~\cite{BEN1,BEN2,BL} to further explore the subject, bringing novel results. For instance, in Ref.~\cite{BEN1} it was shown that the same scalar field theory may describe distinct quantum
mechanical problems, and in Ref.~\cite{BEN2} it was shown that the reconstruction of a field theory model is not unique, that is, we can reconstruct distinct field theories from the very same quantum mechanical potential. These results are of current interest and more recently yet, the investigation in \cite{BL} offered the possibility of constructing distinct field theory models with the very same stability potential, thus leading us to the same quantum mechanical potential.

The current work is aimed to enlarge and deepen the results obtained in Refs.~\cite{BEN1,BEN2,BL}. In this sense, we use the deformation procedure \cite{DP} to obtain distinct field theories associated to the same quantum mechanical potential. For this purpose, we suggest different models related by the deformation procedure, which allows to write a constraint that ensures equality between the stability potentials of the models. The simplest deformation function which satisfies such constraint  permits to study several examples having the feature indicated above. The procedure adopted here works as follows: given a starting model,  the deformation is applied to provide other models that present the same stability potentials of the first one. The starting model may have symmetric or asymmetric topological sectors, and in the case of asymmetric sectors, the deformation procedure gives rise to a new sector, with modified stability potential, besides the ones having the same stability potential, as we show below.

The  examples studied in this paper highlight polynomial and hyperbolic potentials. These kinds of models are useful to investigate defect structures \cite{Guilarte,Lima},  braneworld scenarios with a single extra dimension of infinite extent \cite{brane}, kink-antikink pairs production in collisions of particle-like states \cite{Romanczukiewicz}, successions of phase transitions in field theories \cite{Khare}, and  can also be employed to explore aspects of integrability \cite{Klotzek, Aguirre, Bertini}.

There are other motivations to study kinks, since they may be used to model one-dimensional structures that appear in magnetic materials at the nanometric scale. For instance, in \cite{A1} the authors have experimentally shown that the kinklike profile may change when it is investigated on constrained geometries. Also, in \cite{A2} it was experimentally proven that electric pulses may contribute to change the polarity of a domain defect. More recently, authors have studied kinklike structures in interaction with elastic waves, as in \cite{ew1}, where magnetic bubbles in a bismuth-substituted iron garnet was considered, and also in \cite{ew2}, where a theoretical investigation suggests how to understand the interaction of elastic waves with domain walls.

In this work we concentrate on the construction of kinklike models that engender the same stability potential. In the next section, we review some results about the deformation procedure in systems involving a single real scalar field \cite{DP}. In Sec.~\ref{sec2} one shows how to find deformation functions that give rise to distinct field theories having the same stability potential. In Sec.~\ref{sec3}, we present illustrations obtained by applications of a deformation function to several models with polynomial and hyperbolic interactions. We finish the work adding our comments and conclusions in Sec.~\ref{sec4}.

\section{Generalities} \label{sec1}
 
Let us start briefly revising the deformation procedure introduced in Ref.~{\cite{DP}} for real scalar fields in $(1,1)$ spacetime dimensions. We will use natural units, with $\hbar=1=c$, and shift the field, spacetime coordinates and all the parameters that identify the model to work with dimensionless variables. We take two models defined by the Lagrangian densities
\bens
{{\cal L}}&=&\frac12\partial_\mu\chi\partial^\mu\chi-{U}(\chi)\slabel{1a}
\\
{\cal L}_d&=&\frac12\partial_\mu\phi\partial^\mu\phi-{V}(\phi)\slabel{1b}
\eens
where $\chi$ and $\phi$ are real scalar fields, and ${U}(\chi)$ and $V(\phi)$ are the corresponding potentials. We introduce a deformation function $f=f(\phi)$, from which we link the model \eqref{1a} with the deformed model (\ref{1b}) by relating both potentials ${U}(\chi)$ and $V(\phi)$ in the very specific form
\be\label{vdef}
{V}(\phi)=\frac{{U}(\chi\to f(\phi))}{(df/d\phi)^2}\,.
\ee
This allows to show that under some specific conditions, if the starting model supports the static solution $\chi(x)$, then the deformed model supports the static solution given by $\phi(x)=f^{-1}\lt(\chi(x)\rt)$, where $f^{-1}$ stands for the inverse of the function $f$. The demonstration of this property was already given in Ref.~{\cite{DP}}. 

The deformation procedure is an approach that aims to build localized solutions, so it starts with a model that supports localized solution and help us to find another model, which also support localized solution, constructed with the use of the localized solution of the starting model and the inverse of the deformation function. It is similar but not a field redefinition, since in a field redefinition one has to change the field everywhere in the Lagrangian density, while in the deformation procedure one just change the potential according to the recipe shown in \eqref{vdef}. It has been explored before in the works in Ref.~\cite{DP}, and in several other more recent investigations, in particular in \cite{Lima}, to describe kinklike solutions in models with hyperbolic interactions, and in \cite{D2}, where it is used to describe the presence of kinks in models of the Dirac-Born-Infeld type.

In order to illustrate how the deformation works, let us consider the $\chi^4$ model which will be further explored below and is defined by \eqref{v1}. In this case, the deformation function $f(\phi)=\sin(\phi)$ and the prescription \eqref{vdef} lead us to the potential
\be 
V(\phi)=\frac12 \cos^2(\phi),
\ee
which defines the sine-Gordon model. As one knows, $\chi(x)=\tanh(x)$ is localized solution of the $\chi^4$ model, so $\phi(x)=\arcsin(\tanh(x))$ is localized solution of the sine-Gordon model, as can be verified straightforwardly.

To evaluate linear stability of the static scalar field solutions of theory \eqref{1a}, we assume small fluctuations ${\eta}_n (x)$ around the classical field ${\chi}(x)$
\be
{\chi}(x,t)={\chi}(x)+\sum_{n}{\eta}_{n}(x)\cos({\omega}_n t)\,.
\ee
Substituting this into the time-dependent equation of motion
\be
\frac{\partial^2\chi}{\partial t^2}-\frac{\partial^2\chi}{\partial x^2}+\frac{d{U}}{d\chi}=0\,,
\ee
we obtain the Sch\"odinger-like equation
\be\label{sch}
\left(-\frac{d^2}{dx^2}+{u}(x)\right){\eta}_n (x)={\omega}_n^2\;{\eta}_n (x)\,,
\ee
where ${u}(x)$ is the stability potential, a quantum mechanical potential given by
\be\label{Ut}
{u}(x)=\left.\frac{d^2{U}}{d\chi^2}\right|_{\chi={\chi}(x)}\,.
\ee
The equation (\ref{sch}) has at least one bound state, the bosonic zero mode which is present due to translational invariance. It is given by
\be
{\eta}_0 (x)={{\cal N}}\, \frac{d{\chi}(x)}{dx}\,,
\ee
where ${{\cal N}}$ is a normalization constant.

Considering the situation in which the potential can be written in the alternative form ${U}(\chi)=(1/2){W}_{\chi}^{2}$, where ${W}={W}(\chi)$ and
${W}_{\chi}=d{W}/d\chi$, one can write the energy density in the form
\be
{\epsilon}(x)= \left(\frac{d{\chi}(x)}{dx}\right)^2,
\ee
and the Sch\"odinger-like equation \eqref{sch} can be factorized into ${A}^\dag {A}{\eta}_n (x)={\omega}_n^2{\eta}_n (x)$, where
\be
{A}^\dag=-\frac{d}{dx} - W_{\chi\chi}\,\,\,\,\, \mbox{and} \,\,\,\,\,
{A}=\frac{d}{dx} -W_{\chi\chi}. \nonumber
\ee
This leads to eigenvalues ${\omega}_n^2$ that are positive defined, thus implying that the static solution of the starting field
theory \eqref{1a} is linearly stable. 

In the same way, a similar investigation of the deformed model leads to a similar result, that the static solution is also linearly stable. Here the stability potential has the form
\be\label{U} 
v(x)=\left.\frac{d^2{V}}{d\phi^2}\right|_{\phi(x)=f^{-1}\lt(\chi(x)\rt)}\,,
\ee
the zero mode is given by
\be
\xi_0 (x)={\cal N}\, \frac{d{\phi}(x)}{dx},
\ee
and the energy density is
\be
\rho(x)= \left(\frac{d{\phi}(x)}{dx}\right)^2.
\ee

Note that for the starting model, one uses $\chi,U,\eta,u,$ and $\epsilon$ to represent the field, potential, fluctuation, stability potential and energy density, and for the deformed model one uses $\phi,V,\xi,v$ and $\rho$ to represent the same quantities, respectively.

\section{Procedure} \label{sec2}

The goal here is to introduce distinct field theory models that generate the same stability potential.
Assuming that $\chi=f(\phi)$ and deriving the equation \eqref{vdef} with respect to the variable $\phi$, we have
\ben
&&v(x)=u(x)
-\left.\left(\frac{3f_{\phi\phi}}{f_{\phi}^2}\frac{d{U}}{d\chi}\right)\right|_{\phi=f^{-1}(\chi)}\!\!
+\left.\left(\frac{6f_{\phi\phi}^2}{f_{\phi}^4}{U}-\frac{2f_{\phi\phi\phi}}{f_{\phi}^3}{U}\right)\right|_{\phi=f^{-1}(\chi)}\,,
\een
where $f_{\phi}={df}/{d\phi}$, etc. 

\begin{figure}[ht]
\includegraphics[{height=3cm,width=7cm}]{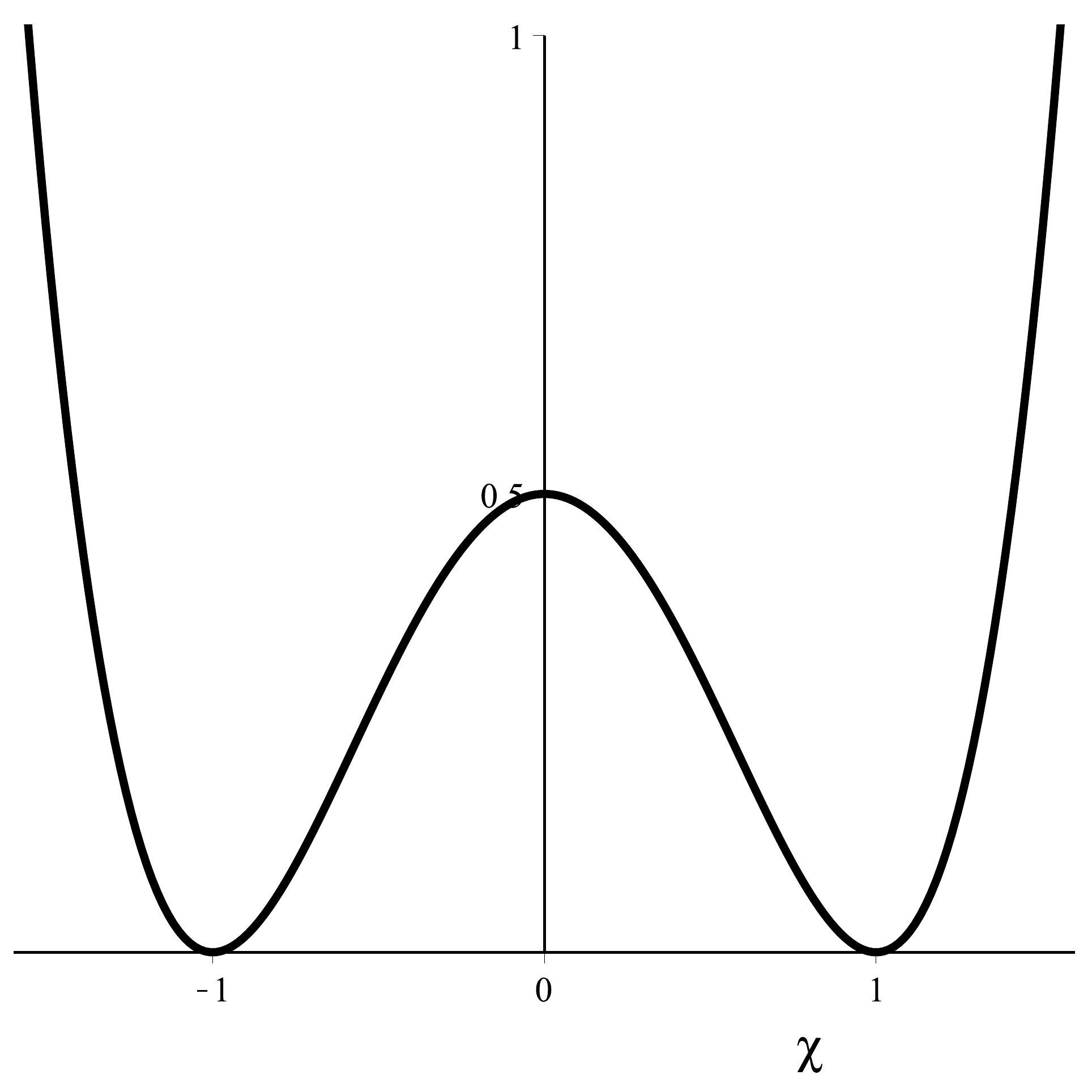}
\caption{The potential \eqref{v1} as a function of the field $\chi$.}\label{fig1}
\end{figure}

\begin{figure}[ht]
\includegraphics[{height=3cm,width=7cm}]{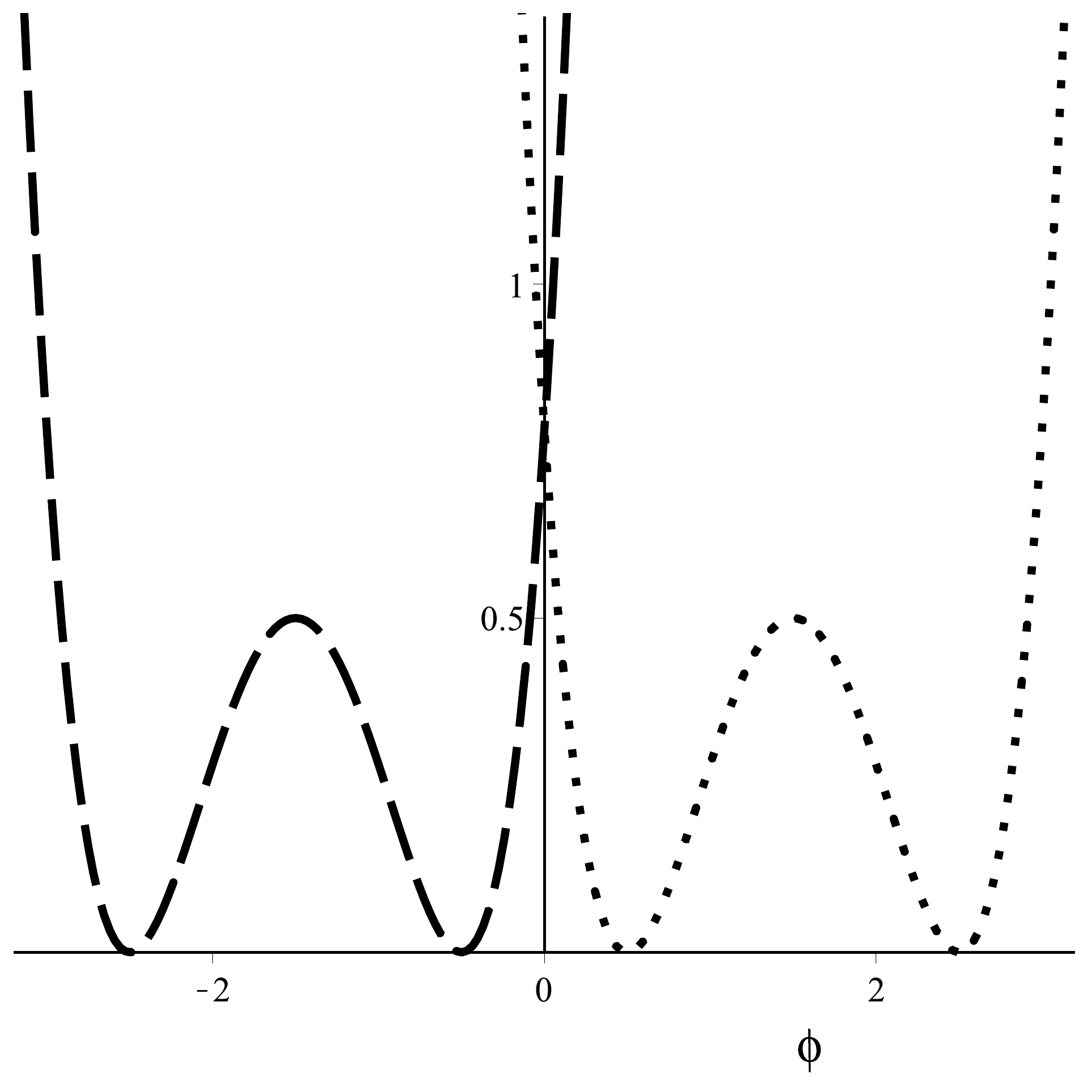}
\includegraphics[{height=3cm,width=7cm}]{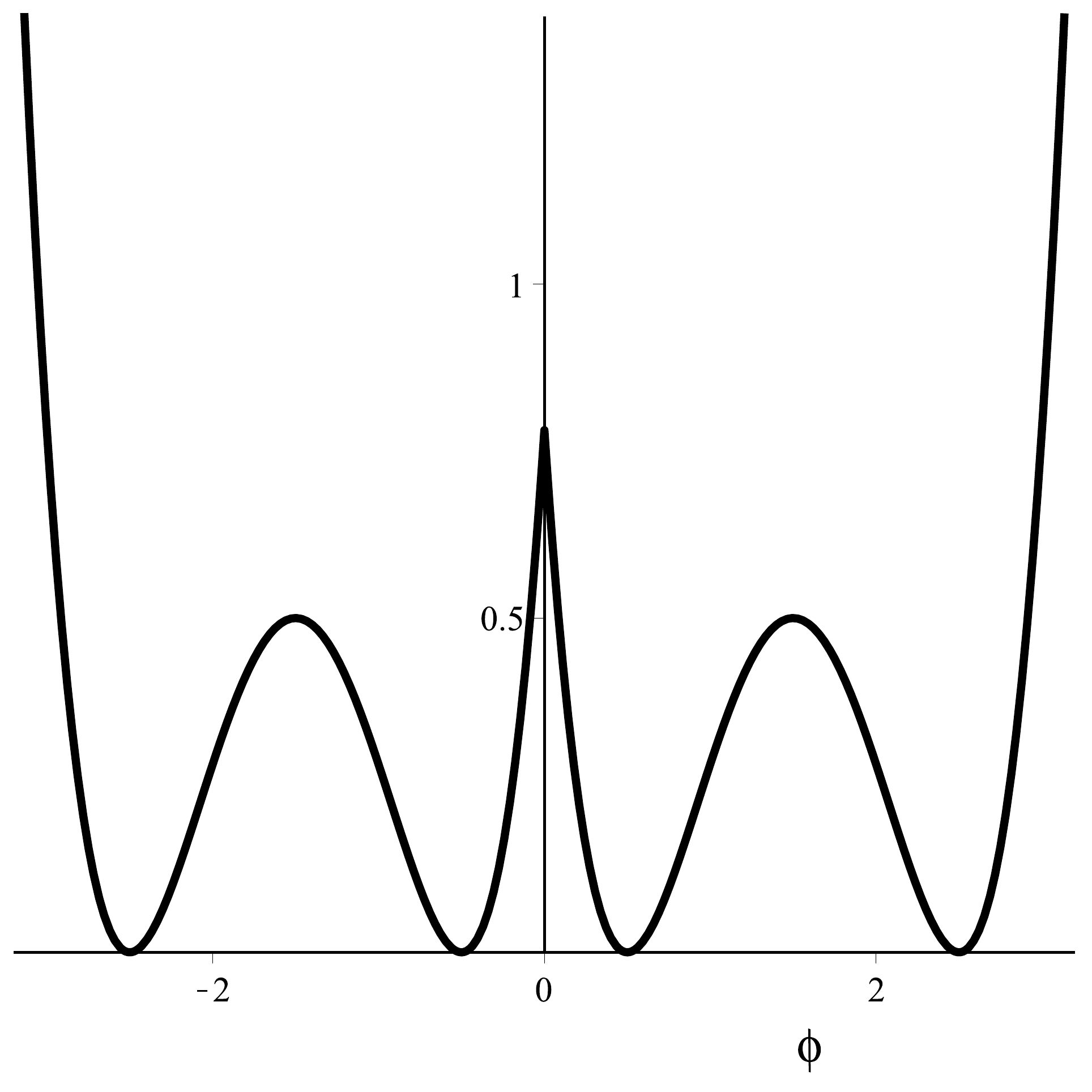}
\caption{In the left panel, one displays the potentials \eqref{v11a} (dashed line) and \eqref{v11b}  (dotted line). In the right panel, one depicts the potential \eqref{v2} (solid line). We take $c=3/2$.}\label{fig1ab}
\end{figure}

Then, in order to have the same stability potential for both theories \eqref{1a} and \eqref{1b}, that is,  for ${u}(x) = v(x)$, the deformation function $f(\phi)$  must satisfy the constraint given by
\be
f_{\phi}f_{\phi\phi\phi}-3f_{\phi\phi}^2+\frac{3}{2}f_{\phi}^2f_{\phi\phi}\frac{1}{{U}}\frac{d{U}}{d\chi}=0.
\ee
For any physically acceptable initial potential ${U}(\chi)$, the restriction above can be solved by deformations of the form
$f(\phi)=a+b\,\phi,$ with $a,b$ as real constants. Let us consider, for instance, the two deformation functions below
\bens
\label{fs}
f_1(\phi)&=&\phi+c\,,\\
f_2(\phi)&=&\phi-c\,,
\eens 
where $c>0$ is a real constant. As a first step, to better understand how this procedure works, we apply these deformations to the $\chi^4$ potential given by 
\be\label{v1}
{U}(\chi)=\frac12(1-\chi^2)^2\,,
\ee
which  has one topological sector, as illustrated in Fig.~\ref{fig1}, characterized by the two minima $\chi_{\min}=\pm 1$, the static solutions
$\chi(x) =\pm \tanh(x)$, and the energy density ${\epsilon}(x)=\sech^4(x)$. The functions \eqref{fs} provide two deformed potentials given  respectively by
\bens\label{v11}
\slabel{v11a}
{V}_1&=&\frac12\left(1-c^2-2c\phi-\phi^2\right)^2\,, \\
\slabel{v11b}
{V}_2&=&\frac12\left(1-c^2+2c\phi-\phi^2\right)^2\,.
\eens
It is found that these potentials produce displacements in the model $\chi^4$, as can be seen in the left panel of Fig.~\ref{fig1ab}, where the first one causes a shift to the left, and the second one a shift to the right.
Furthermore, these two potentials \eqref{v11} can be {\it sewed} together to become a single one, given by
\be\label{v2}
V(\phi)=\frac12\left(1-c^2+2c|\phi|-\phi^2\right)^2\,,
\ee
which is obtained deforming \eqref{v1} by  means of the function 
\be\label{def1}
f(\phi)=c-|\phi|\,.
\ee
Each value of $c$ determines the point where the two deformed potentials \eqref{v11}  are cut to form the new potential \eqref{v2}, as illustrated in the right panel in Fig.~\ref{fig1ab}. 

As we are going to show, in order to obtain the deformed potential as a smooth function of the scalar field, one has to choose for $c$ the extreme values of the potential we started with, remembering that we choose $c>0$ to obtain different potentials. The deformation function \eqref{def1} is then crucial to write distinct field theory models, having the very same stability potential and so leading us to the same quantum mechanics. This was firstly pointed out in Ref.~\cite{BL}, and here we show how this applies to several distinct models described by a single real scalar field with standard dynamics. 

\section{Applications} \label{sec3}
Since the procedure for finding distinct field theories having similar quantum mechanical potential was just built in the previous subsection, let us now consider different potentials that engender polynomial and non-polynomial self-interactions, which are related by a deformation function and supports the indicated feature for the stability potential. 

\subsection{The case of polynomial potentials}

\subsubsection{Example 1}

\begin{figure}[ht]
\includegraphics[{height=3cm,width=7cm}]{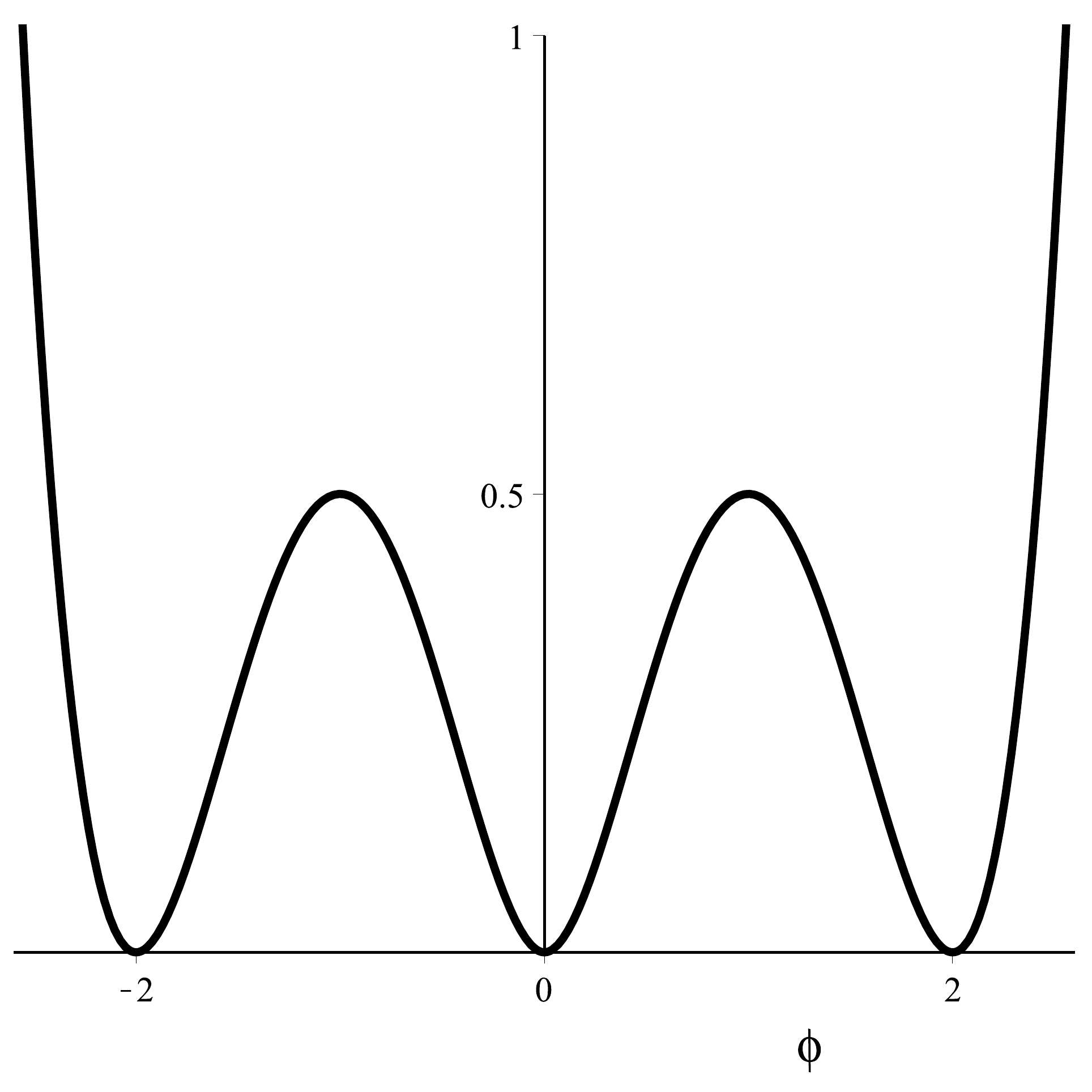}
\caption{The potential \eqref{v21} as a function of the field $\phi$.}
\label{fig2}
\end{figure}

\begin{figure}[ht]
\includegraphics[{height=3cm,width=7cm}]{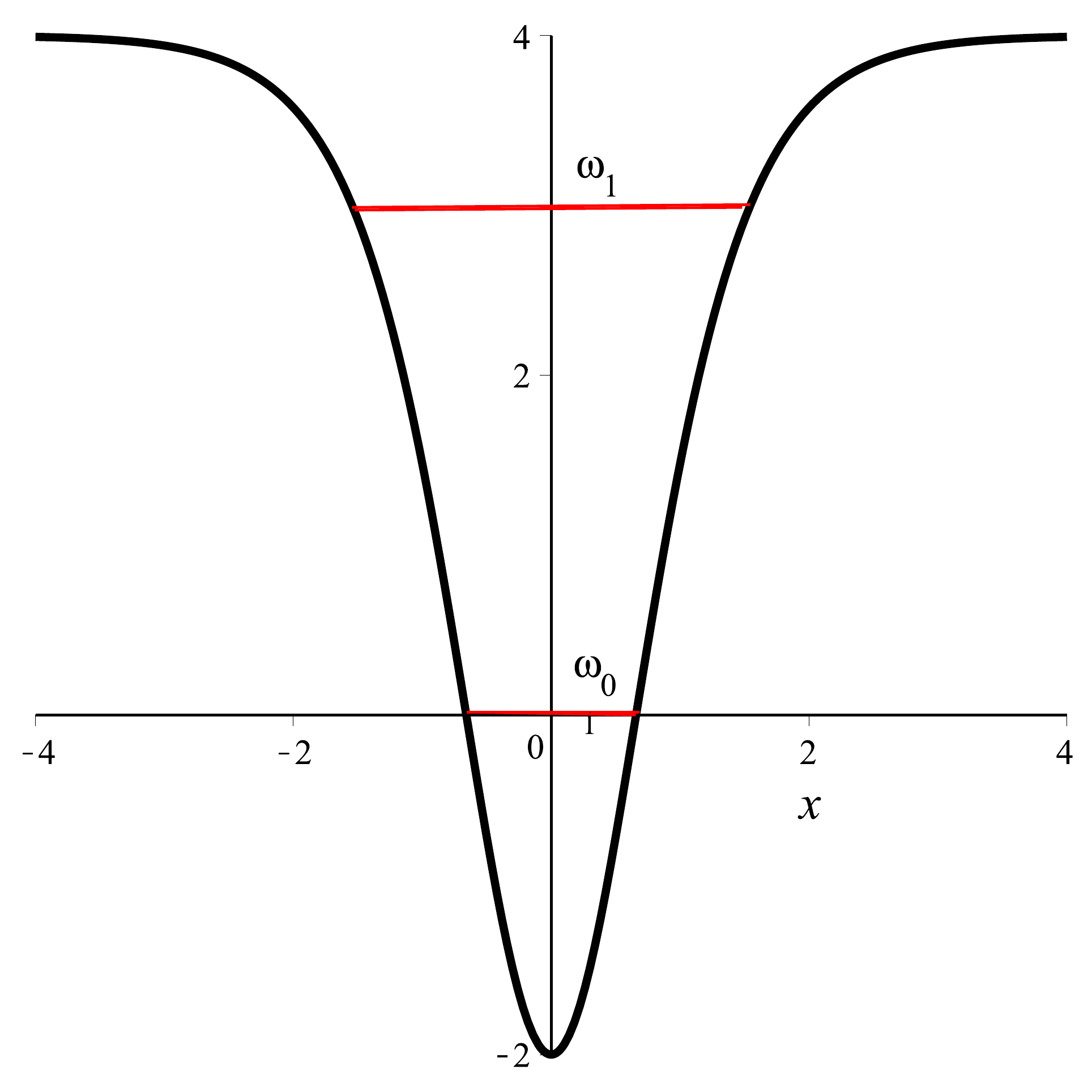}
\caption{The stability potential \eqref{u1} and the two bound states, represented by the two red (thinner) horizontal lines.}
\label{figu2}
\end{figure}

Let us further explore the properties of the potential \eqref{v2}, which was obtained considering the deformation function \eqref{def1} acting on the $\chi^4$ theory described by the potential \eqref{v1}. The new potential is a smooth function of $\phi$ for $c=1$, which corresponds to a minimum of the starting potential \eqref{v1}. In this case, the expression \eqref{v2} becomes
\be\label{v21}
V(\phi)=\frac12\phi^2(2-|\phi|)^2\,,
\ee
which has two topological sectors, as illustrated in Fig.~\ref{fig2}. It has the static solutions $\phi(x)=\pm\tanh(x)\pm1$ with the corresponding  energy density $\rho(x)=\sech^4(x)$. From  Eq.~\eqref{U}, we obtain the same stability potential for both topological sectors; see also Ref.~\cite{BEN2}. It is given by
\be\label{u1}
v(x)=4- 6\,\sech^2(x)\,,
\ee
which is of the modified P\"oschl-Teller type \cite{MF}, equivalent to the one obtained by the $\chi^4$ model, presenting two bound states, the zero mode with eigenvalue $\omega_0^2 = 0$ and the excited state with $\omega_1^2 = 3$, as illustrated in Fig.~\ref{figu2}. The zero mode has the form 
$\eta_0 (x)={\cal N}\,\sech^2(x),$ where ${\cal N}$ is the normalization constant. 

\subsubsection{Example 2}

\begin{figure}[ht]
\includegraphics[{height=3cm,width=7cm}]{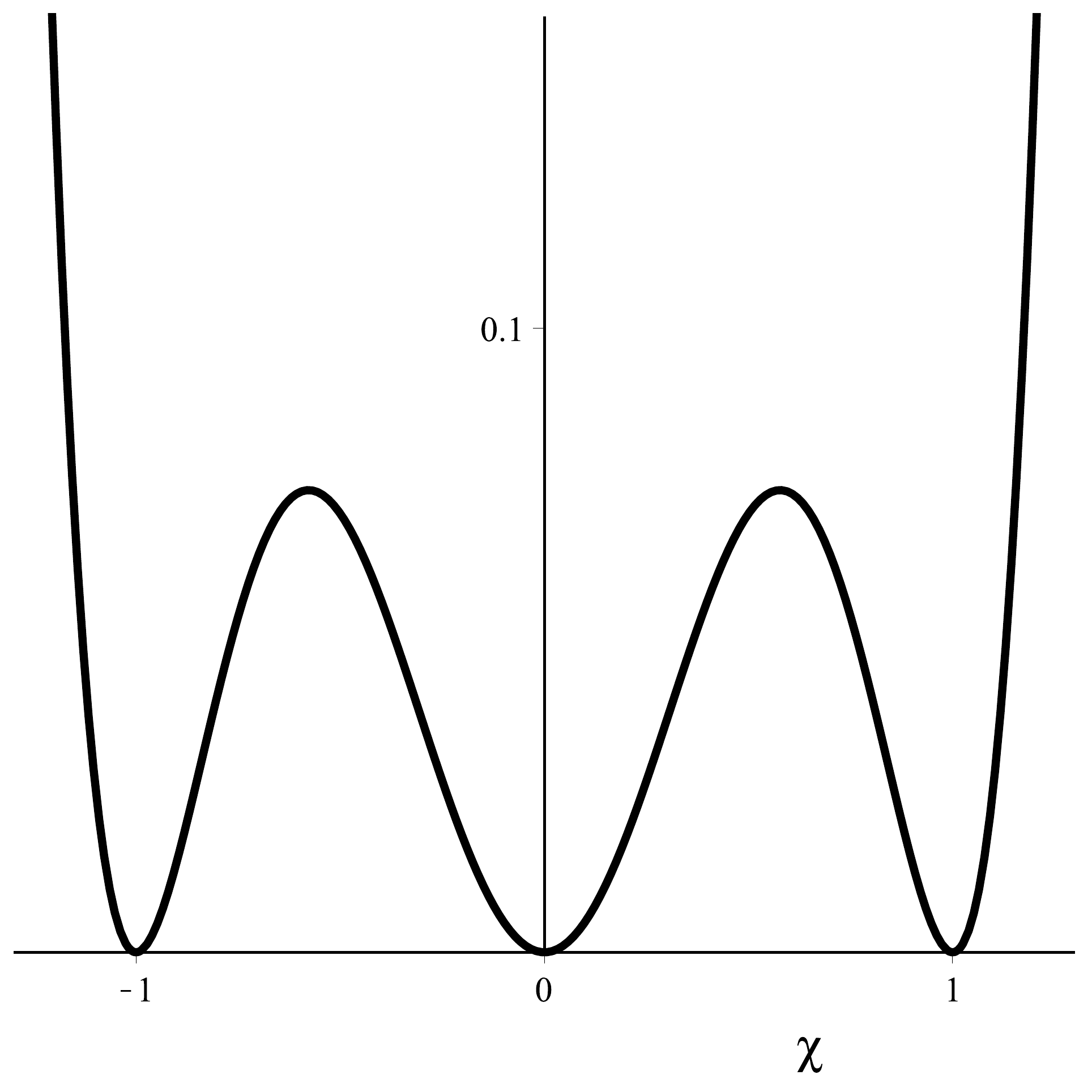}
\caption{The $\chi^6$ potential \eqref{v5} as a function of the field $\chi$.}\label{fig3}
\end{figure}

We now take the $\chi^6$ model as the starting field theory,
\be\label{v5}
{U}(\chi)=\frac12\chi^2(1-\chi^2)^2\,,
\ee
which has the minima $\chi_{\min}=0,\pm1$ and maxima at $\chi_{\max}=\pm 1/\sqrt{3}$, characterizing two topological sectors, as illustrated in Fig.~\ref{fig3}, whose kink solutions are given by
\bens\label{sol51}
\chi_1(x)&=&-\sqrt{\frac{1-\tanh(x)}{2}}\,, \\
\chi_2(x)&=&\sqrt{\frac{1+\tanh(x)}{2}}\,,
\eens
and are associated to the left and right sectors of the $\chi^6$ model, respectively. Although they have the same energy, the energy densities corresponding to these solutions are different, given by 
\bens\label{ent1}
\slabel{ent1a}
{\epsilon}_1(x)&=&\frac{1}{8}\sech^2(x)\lt(1+\tanh(x)\rt),\\
\slabel{ent1b}
{\epsilon}_2(x)&=&\frac{1}{8}\sech^2(x)\lt(1-\tanh(x)\rt).
\eens

\begin{figure}[ht]
\includegraphics[{height=3cm,width=7cm}]{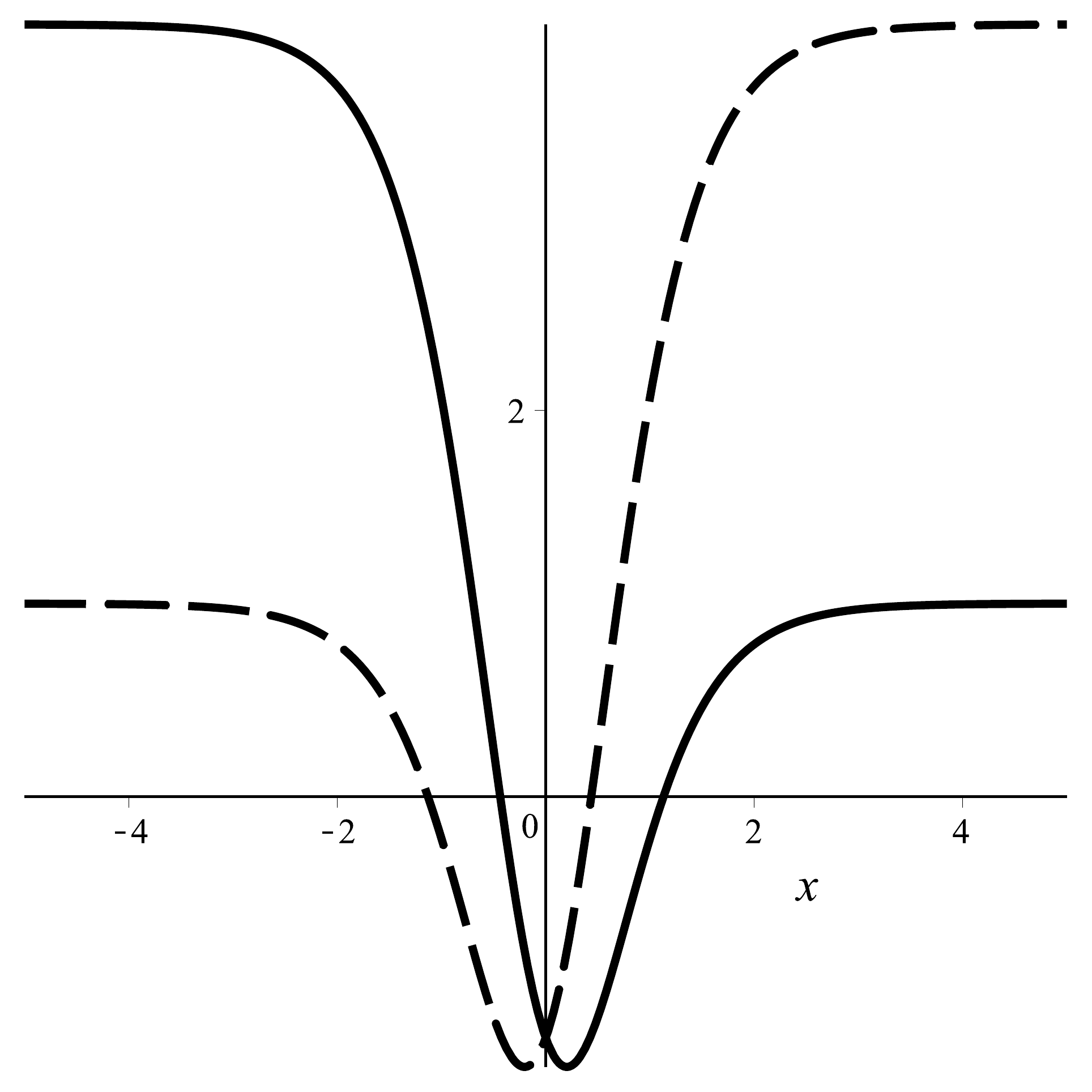}
\caption{The stability potentials \eqref{u51} (solid line) and \eqref{u52} (dashed line).}\label{figu3}
\end{figure}

Replacing the potential \eqref{v5} and the solutions \eqref{sol51} into Eq.~\eqref{Ut}, we obtain the two stability potentials
\bens
{u}_1(x)&=&\frac52-\frac32\tanh(x)-\frac{15}{4}\,{\rm sech}^2(x)\,,\slabel{u51}\\
{u}_2(x)&=&\frac52+\frac32\tanh(x)-\frac{15}{4}\,{\rm sech}^2(x)\,,\slabel{u52}
\eens
which have only one bound state related to the eigenvalue $\omega^2_0=0$, which represents the zero mode \cite{lohe}. These potentials are illustrated in Fig.~\ref{figu3}.

We now apply the deformation function \eqref{def1} to the potential \eqref{v5}, as shown in Eq.~\eqref{vdef}, to obtain the deformed model 
\be\label{v51}
V(\phi)=\frac12(c-|\phi|)^2\left(1-c^2+2c|\phi|-\phi^2\right)^2\,,
\ee
which is a smooth function of $\phi$ for $c=1$ and for $c=1/\sqrt{3}$.

\begin{figure}[ht]
\includegraphics[{height=3cm,width=7cm}]{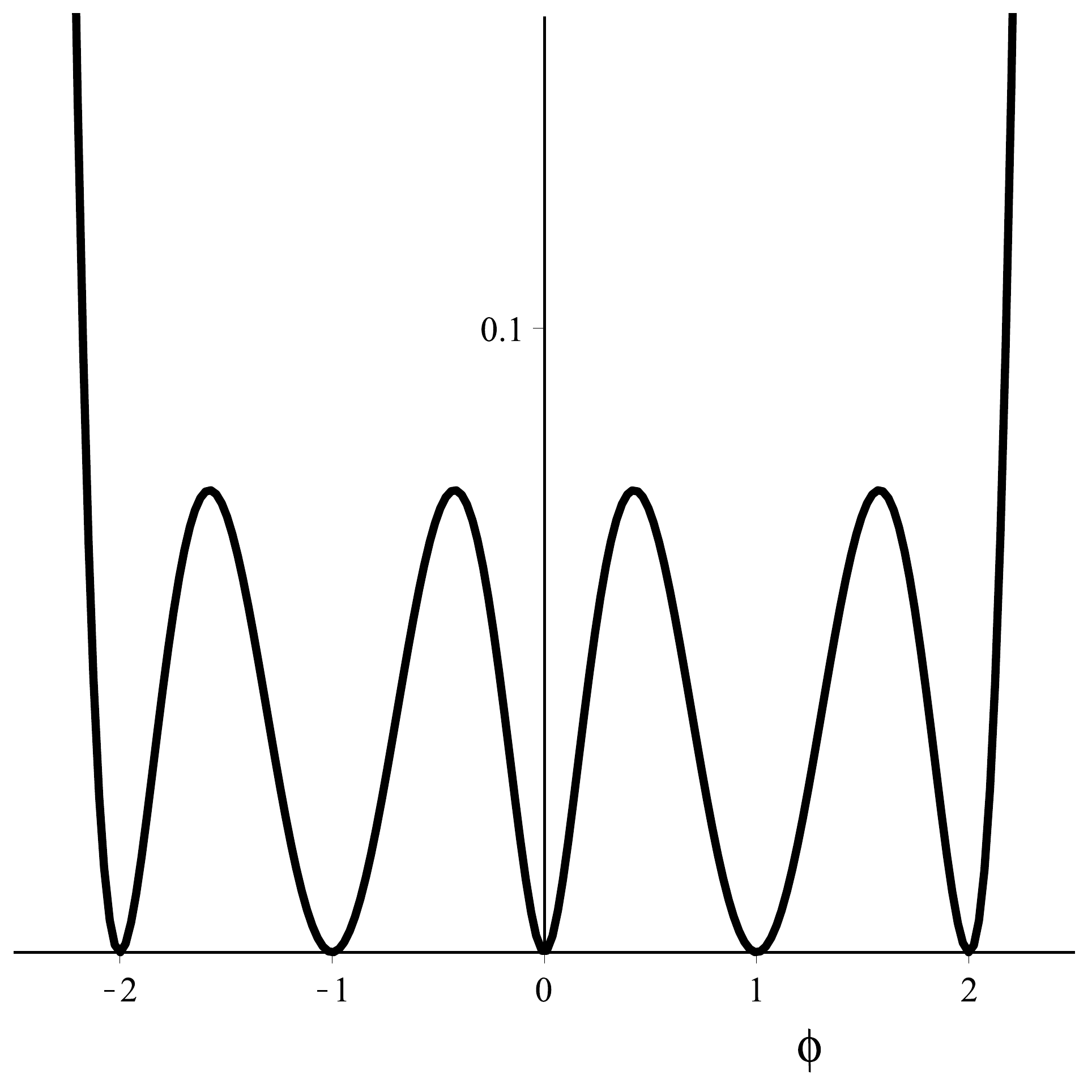}
\caption{The potential \eqref{v6} as a function of the field $\phi$, for $c=1$.}\label{fig4}
\end{figure}

For $c=1$, the potential \eqref{v51} is written as
\be\label{v6}
V(\phi)=\frac12\phi^2(2-|\phi|)^2(1-|\phi|)^2\,,
\ee
that has four topological sectors, as illustrated in Fig.~\ref{fig4}, and its static kink solutions, from left to right, are given respectively by
\bens\slabel{sol61}
\phi_i(x)&=&-1+\chi_i(x)\,,\,\,\, i=(1,2)\,, \\
\phi_j(x)&=&1+\chi_{j-2}(x)\,,\,\,\, j=(3,4)\,,
\eens
with $\chi_k$ given by \eqref{sol51}. There are two distinct energy densities, which are also given by Eqs.~\eqref{ent1}. From  Eq.~\eqref{U}, the two topological sectors related to the solutions $\phi_1(x)$ and $\phi_3(x)$ provide the stability potential \eqref{u51}, and the others two topological sectors related to the solutions $\phi_2(x)$ and $\phi_4(x)$ provide the stability potential \eqref{u52}.

\begin{figure}[ht]
\includegraphics[{height=3cm,width=7cm}]{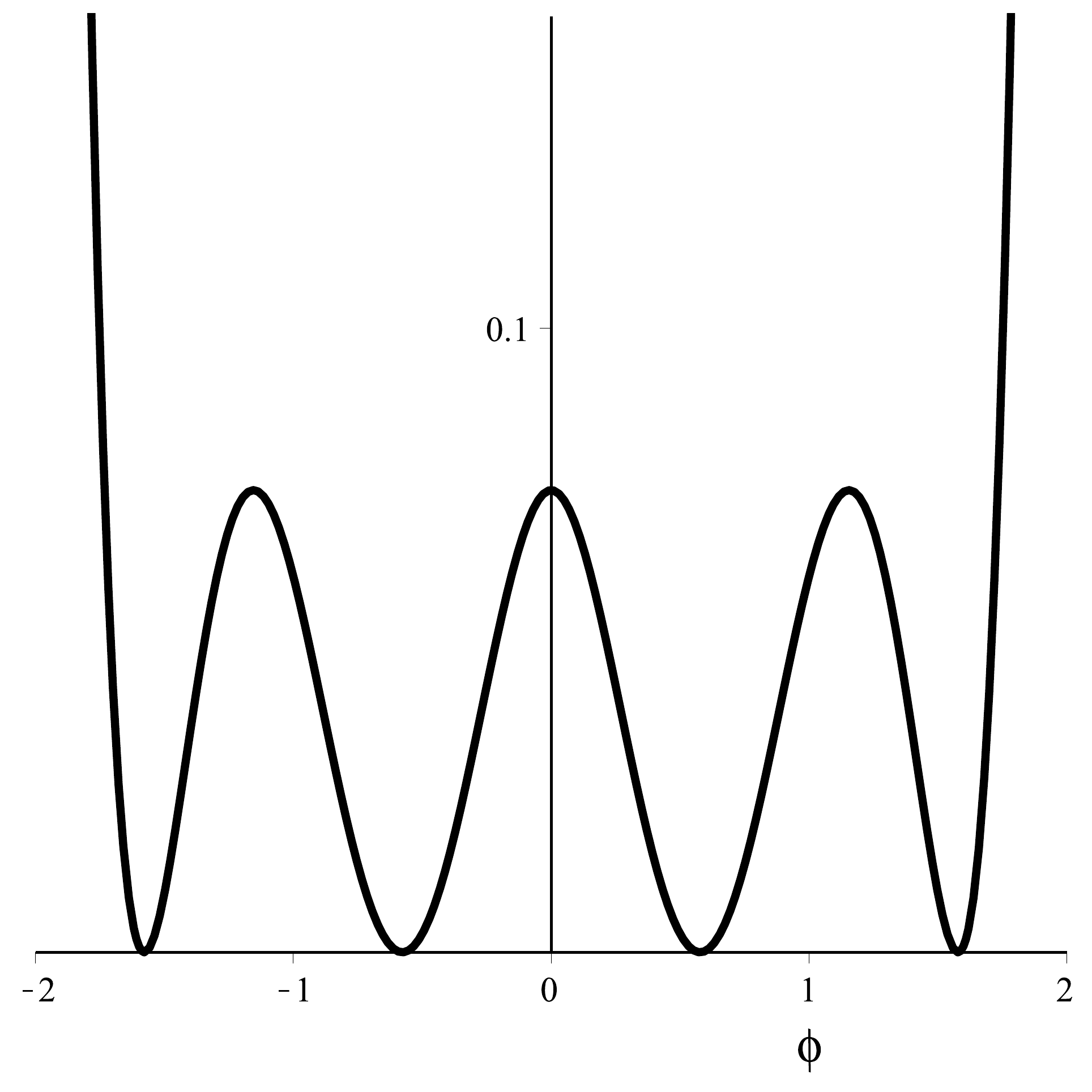}
\caption{The potential \eqref{v6b} as a function of the field $\phi$, for $c=1/\sqrt{3}$.}\label{fig5}
\end{figure}

\begin{figure}[ht]
\includegraphics[{height=3cm,width=7cm}]{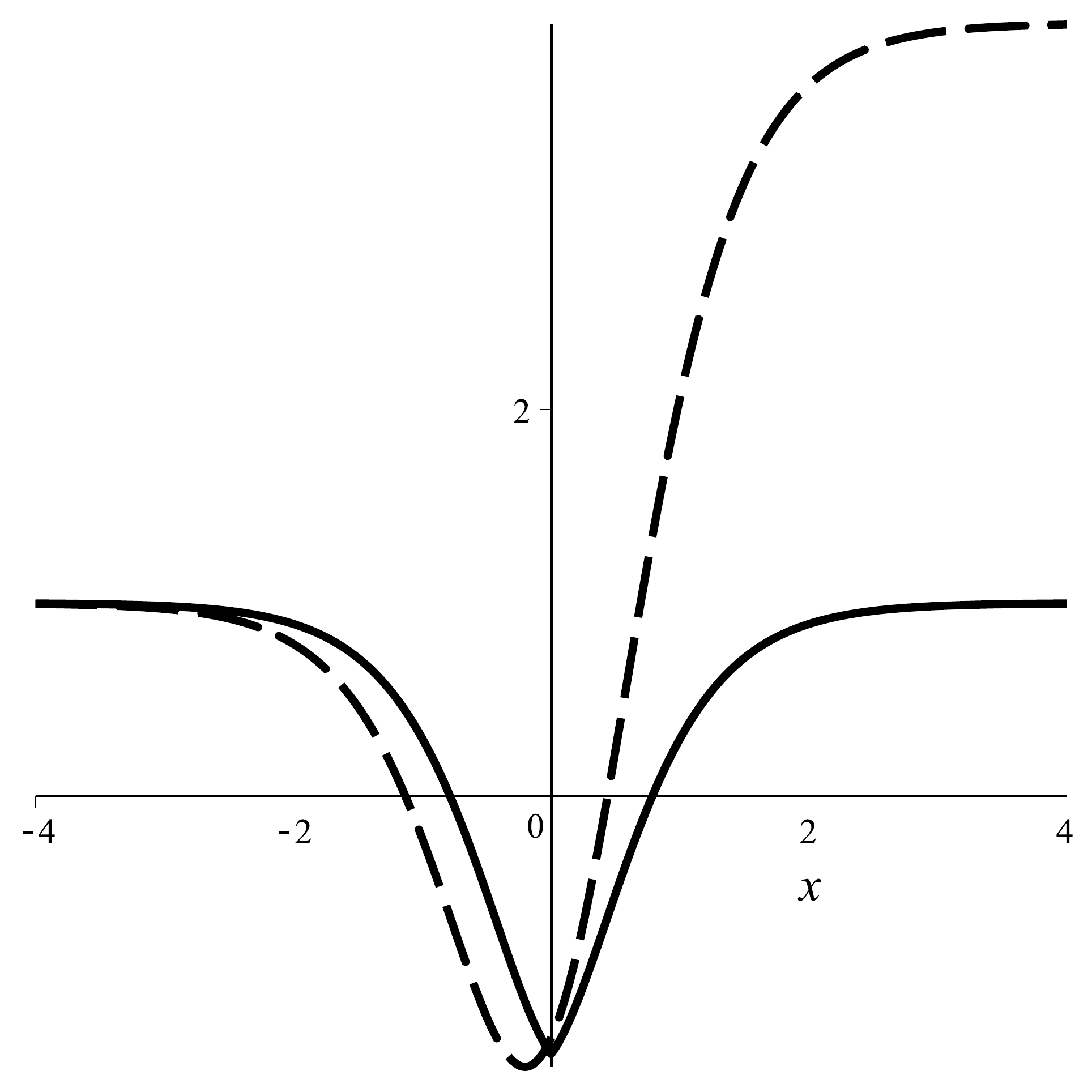}
\caption{The stability potential \eqref{u52} (dashed line) and the new potential \eqref{u7c} (solid line).}\label{figu5}
\end{figure}

For $c=1/\sqrt{3}$, the potential \eqref{v51}  becomes
\be
\label{v6b}
V(\phi)=\frac12\lt(\frac{1}{\sqrt{3}}-|\phi|\rt)^2\left(\frac{2}{3}+\frac{2}{\sqrt{3}}|\phi|-\phi^2\right)^2\,,
\ee
which has three topological sectors, as illustrated in Fig.~\ref{fig5}.
For the two lateral sectors, the kink-like  solutions are 
\bens\slabel{sol71}
\phi_1(x)&=&-\frac{1}{\sqrt{3}}-\sqrt{\frac{1-\tanh(x)}{2}}\,, \\
\slabel{sol72}
\phi_3(x)&=&\frac{1}{\sqrt{3}}+\sqrt{\frac{1+\tanh(x)}{2}}\,.
\eens
They also provide the stability potentials \eqref{u51} and \eqref{u52} and the energy densities \eqref{ent1a} and \eqref{ent1b}, respectively. However, the central topological sector is a new sector, and there one obtains the static solution 
\begin{equation}\label{sol7c}
{\phi}_2(x)=
\left\{
\begin{array}{c}
-\frac{1}{\sqrt{3}}+\sqrt{\dfrac{1+\tanh(x-x_0)}{2}},\;\;\;\;\;x\leq 0\,,\\
\,\,\\
\,\,\,\,\frac{1}{\sqrt{3}}-\sqrt{\dfrac{1-\tanh(x+x_0)}{2}},\;\;\;\;\;x > 0\,,
\end{array}
\right.
\end{equation}
where $x_0={\rm arctanh}(1/3)$ is an adjustment made due to the symmetry of the central sector. 
The energy density corresponding to this sector is
\be
{\rho}_c(x)=
\left\{
\begin{array}{c}
\frac{1}{8}\sech^2(x-x_0)\lt(1-\tanh(x-x_0)\rt), \;\; x\leq 0\,,\\
\,\,\\
\frac{1}{8}\sech^2(x+x_0)\lt(1+\tanh(x+x_0)\rt), \;\; x > 0\,.
\end{array}
\right.
\end{equation}
Also, the stability potential is obtained replacing  \eqref{v6b} and \eqref{sol7c} into  Eq.~\eqref{U},  
\begin{equation}\label{u7c}
{v}_c(x)=
\left\{
\begin{array}{c}
\frac52+\frac32\tanh(x-x_0)-\frac{15}{4}\,{\rm sech}^2(x-x_0),\;x\leq 0\,,
\\\,\,\\
\frac52-\frac32\tanh(x+x_0)-\frac{15}{4}\,{\rm sech}^2(x+x_0),\;x> 0\,,
\end{array}
\right.
\end{equation}
 which is illustrated in Fig.~\ref{figu5} in comparison with the stability potential \eqref{u52}. It has only the zero mode as bound state.

\subsubsection{Example 3}

\begin{figure}[ht]
\includegraphics[{height=3cm,width=7cm}]{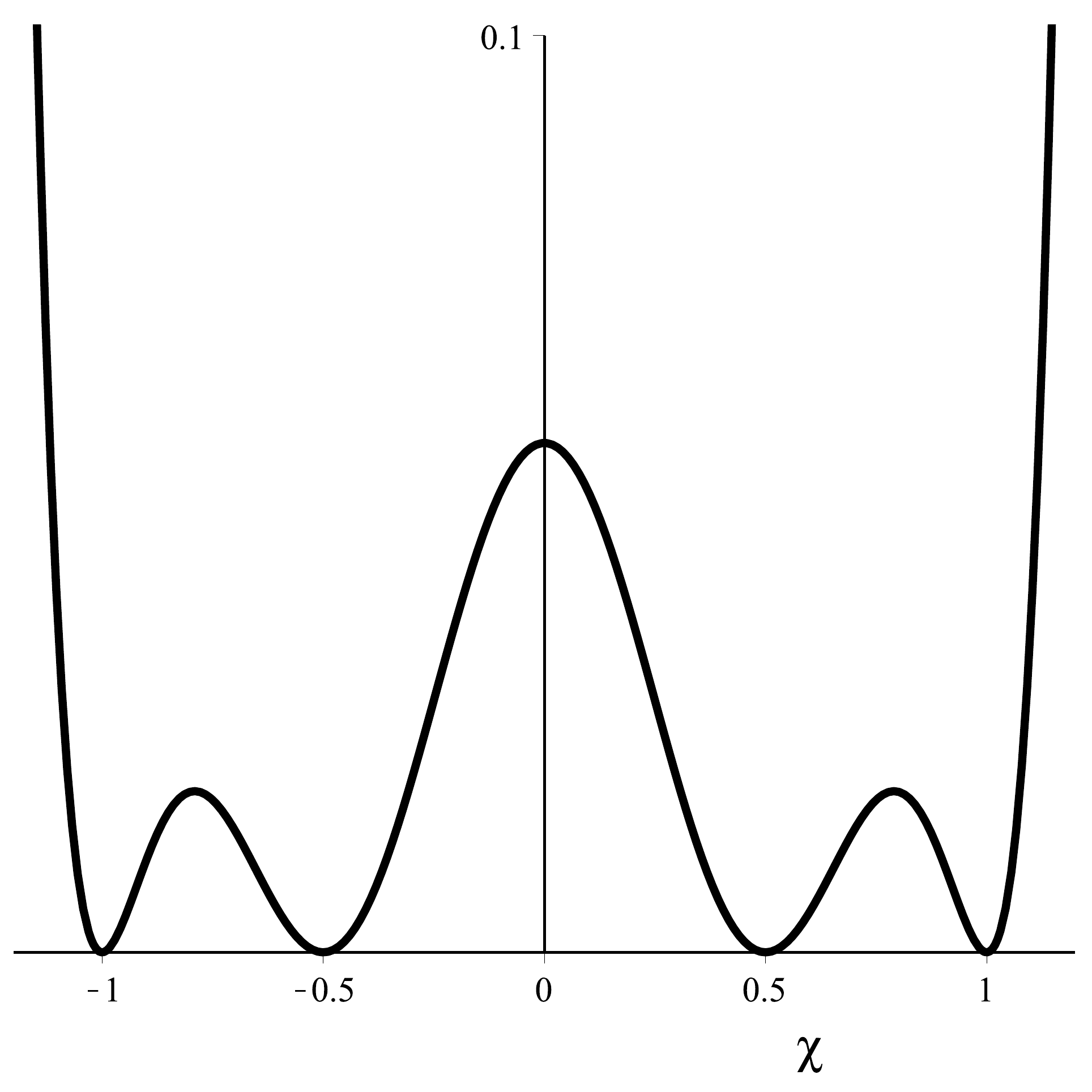}
\caption{The potential \eqref{v8} as a function of the field $\chi$.}\label{fig6}
\end{figure}

We consider the $\chi^8$ potential
\be\label{v8}
{U}(\chi)=\frac89\lt(1-\chi^2\rt)^2\lt(\frac14-\chi^2\rt)^2\,,
\ee
which has four minima at $\chi_{\min}=\pm 1/2, \pm1$ and three maxima at $\chi_{\max}=0,\pm\sqrt{5/8}$, describing three topological sectors as illustrated in Fig.~\ref{fig6}. The static kink solutions are, from left to right, given respectively by
\be\label{sol81}
\chi_n(x)=\cos\left(\frac13\theta(x)+\frac{(3-n)}3\pi\right)\,,\,\,\,\,n=(1,2,3)\,,
\ee
where $\theta(x)=\arccos(\tanh(x))$, so $\theta(x) \in [0,\pi] $. The corresponding energy densities are
\be\label{rhop3}
{\epsilon}_{n}(x)=\frac{1}{9}\sech^2(x)\sin^2\lt(\frac13\theta(x)+\frac{(3-n)}3\pi\rt)\,.
\ee

\begin{figure}[ht]
\includegraphics[{height=3cm,width=7cm}]{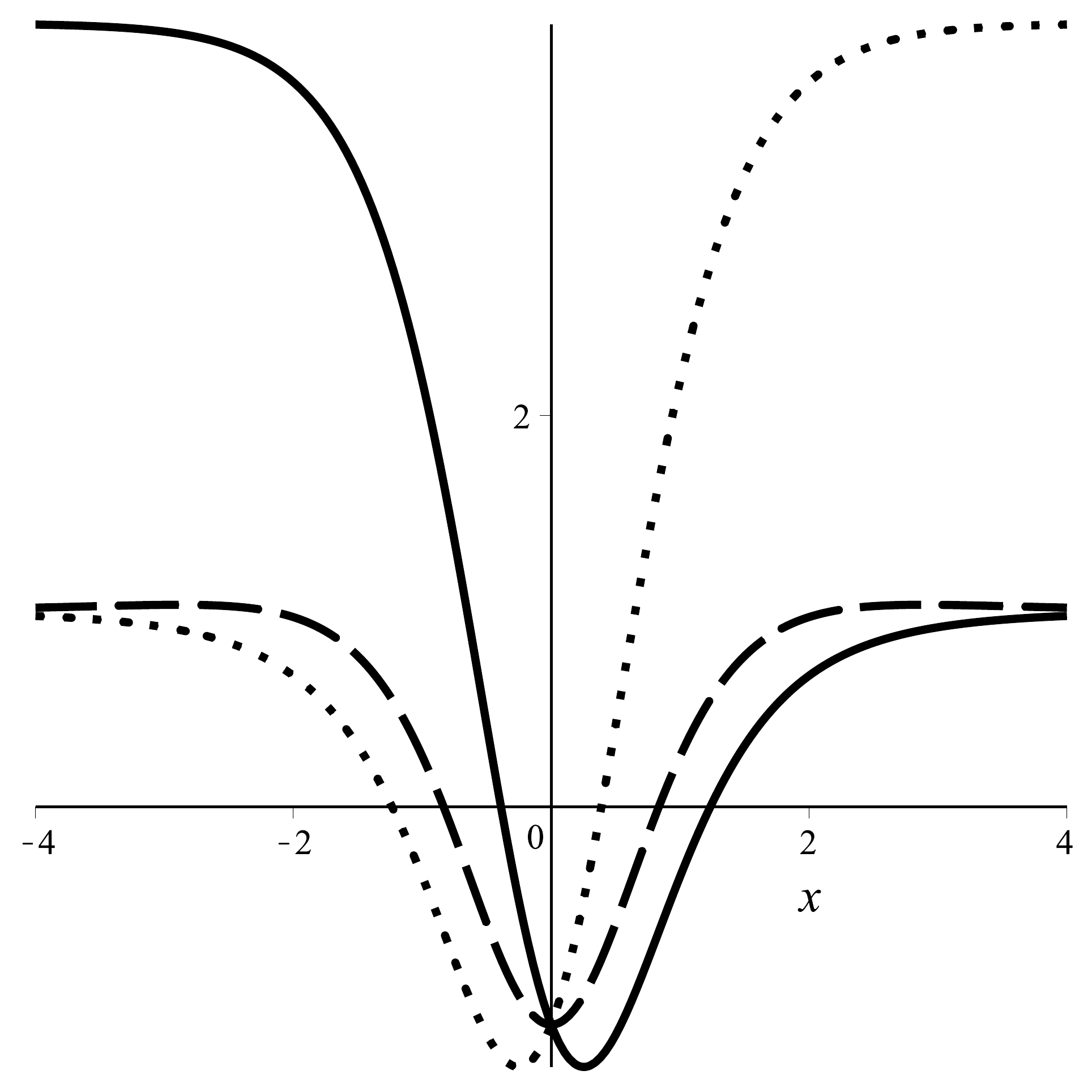}
\caption{The stability potential \eqref{u8}, for $n=1$ (solid line),  $n=2$ (dashed line) and $n=3$ (dotted line).}\label{figu6}
\end{figure}

Replacing Eq.~\eqref{v8} into Eq.~\eqref{Ut} we obtain three stability potentials
\be\label{u8}
{u}_n(x)=-\frac{10}{9}+22\,\chi_n^2(x)-\frac{200}{3}\chi_n^4(x)+\frac{448}{9}\chi_n^6(x)\,,
\ee
where $n=1,2,3$ with $\chi_n(x)$ given by Eqs.~\eqref{sol81}, as illustrated in Fig.~\ref{figu6}.

Here, we consider the deformation function \eqref{def1} applied to the potential \eqref{v8},  then  we obtain the deformed potential given by
\be\label{v81}
V(\phi)=\frac1{18}\left(1-c^2+2c|\phi|-\phi^2\right)^2\left(1-4c^2+8c|\phi|-4\phi^2\right)^2.
\ee 
This model  is a smooth function for $c=1,\sqrt{5/8}$, and $1/2$, as we will consider below.

\begin{figure}[ht]
\includegraphics[{height=3cm,width=7cm}]{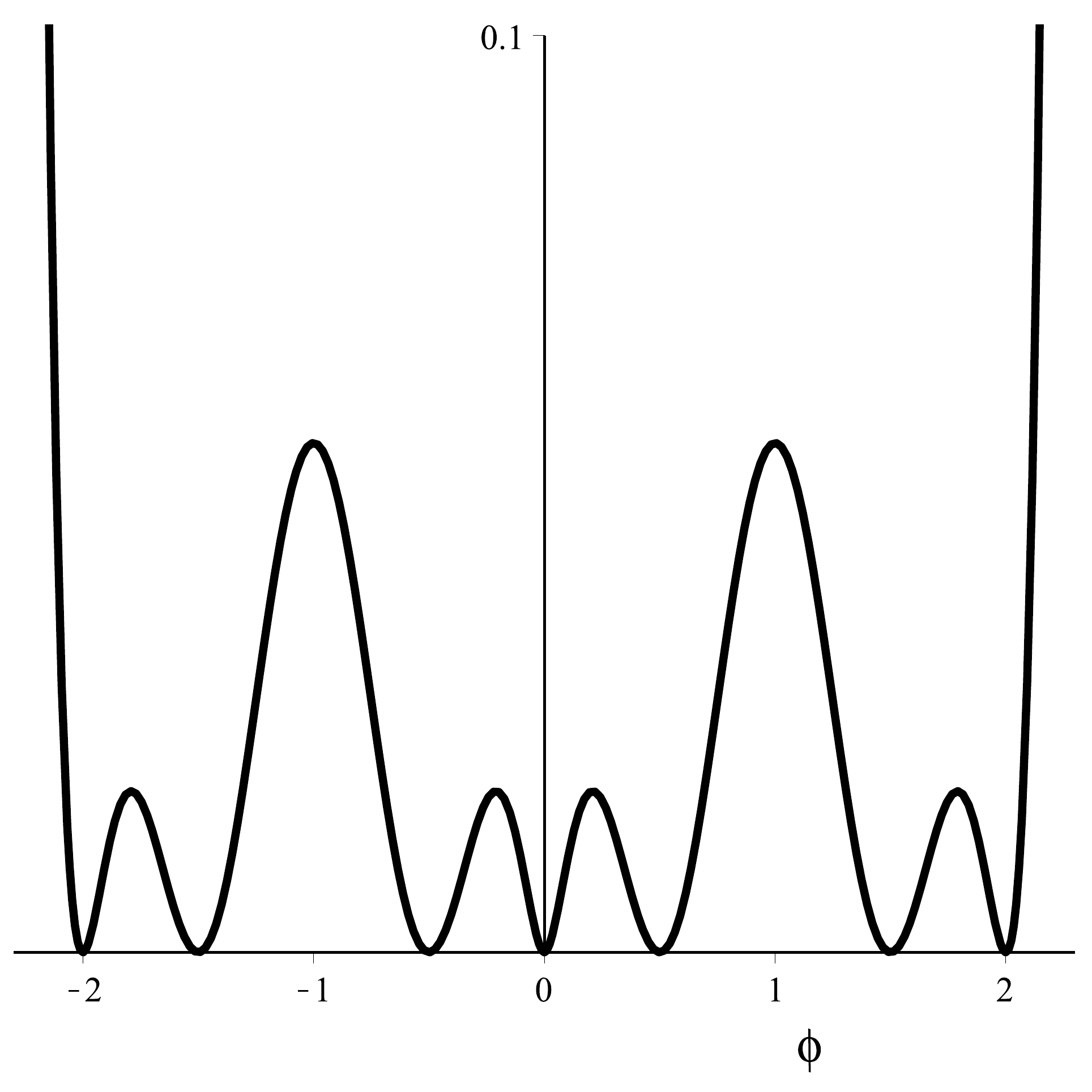}
\caption{The potential \eqref{v9} as a function of the field $\phi$, for $c=1$.}\label{fig7}
\end{figure}

For $c=1$, the potential \eqref{v81} is written as
\be\label{v9}
V(\phi)=\frac1{18}\phi^2(2-|\phi|)^2\left(3+4\phi^2-8|\phi|\right)^2\,,
\ee
which has six topological sectors, as illustrated in Fig.~\ref{fig7}, and its static kink  solutions,  from left to right,  are given in sequence
\bens\label{sol91}
\phi_i(x)&=&-1+\chi_i(x)\,,\,\,\,i=(1,2,3)\,,\\
\phi_j(x)&=&1+\chi_{j-3}(x)\,,\,\,\,j=(4,5,6)\,,
\eens
with $\chi_k(x)$ given by Eqs.~\eqref{sol81}. The corresponding energy densities are equivalent to the ones obtained in Eq.~\eqref{rhop3}. Also, from  Eq.~\eqref{U}, the topological solutions above furnish the stability potentials \eqref{u8} doubly repeated.

\begin{figure}[ht]
\includegraphics[{height=3cm,width=7cm}]{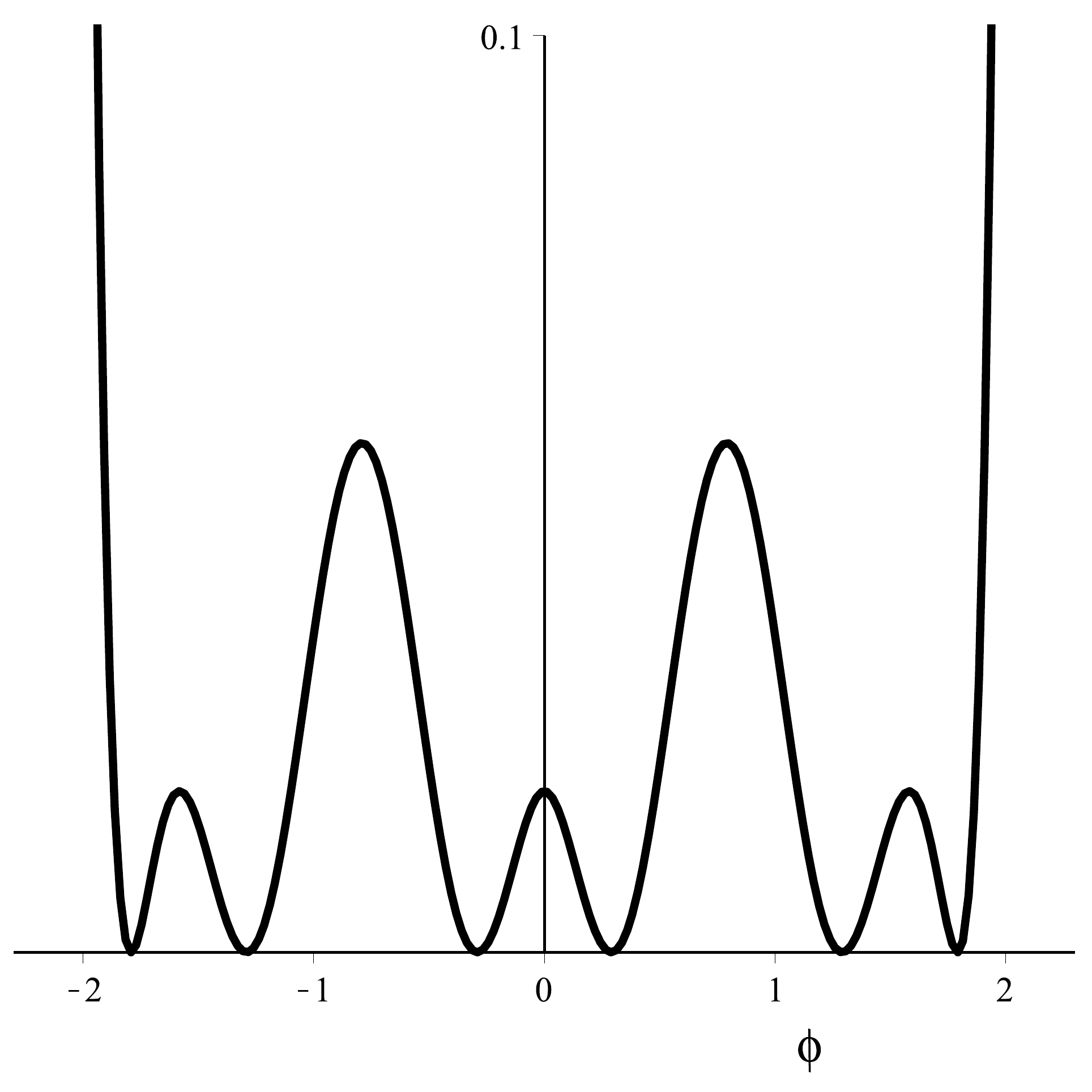}
\caption{The potential \eqref{v95} as a function of the field $\phi$, for $c=\sqrt{5/8}$.}\label{fig8}
\end{figure}

For $c=\sqrt{5/8}$, the potential \eqref{v81} is written as
\be\label{v95}
V(\phi)=\frac1{18}\lt(\frac{3}{8}-\phi^2+\sqrt{\frac52}|\phi|\rt)^2\left(\frac32+4\phi^2-2\sqrt{10}|\phi|\right)^2,
\ee
which has five topological sectors, as illustrated in Fig.~\ref{fig8}, and its static kink  solutions without the central one are
\bens\label{sol95}
\phi_i(x)&=&-\sqrt{\frac58}+\chi_i(x)\,,\,\,\,\, i=(1,2)\,,\\
\phi_j(x)&=&\sqrt{\frac58}+\chi_{j-2}(x)\,,\,\,\,\, j=(4,5)\,,
\eens
with $\chi_k(x)$ given by Eqs.~\eqref{sol81}.
These solutions also provide the energy densities \eqref{rhop3} and the stability potentials \eqref{u8}. For the central topological sector the static kink solution is
\begin{equation}\label{sol75c}
{\phi}_3(x)=
\left\{
\begin{array}{c}
-\sqrt{\dfrac58}+\cos\lt(\dfrac13\theta(x+x_0)\rt),\;\;\;\;\;\;\;x\leq 0,\\
\,\,\\
\sqrt{\dfrac58}+\cos\lt(\dfrac13\theta(x-x_0)+\dfrac{2}{3}\pi\rt),\;x> 0,
\end{array}
\right.
\end{equation}
where $\theta(x-x_0)=\arccos(\tanh(x-x_0))$, $\theta(x+x_0)=\arccos(\tanh(x+x_0))$, and $x_0={\rm arctanh}\lt(\cos\lt(3\arccos(\sqrt{5/8})\rt)\rt)$. The energy density associated to this solution is
\begin{equation}\label{ene3d}
{\rho}_c(x)=
\left\{
\begin{array}{c}
\dfrac{1}{9}\sech^2(x+x_0)\sin^2\lt(\frac13\theta(x+x_0)\rt),\;\;\; \;\;\;\;\;\; x\leq 0,\\
\,\,\\
\dfrac{1}{9}\sech^2(x-x_0)\sin^2\lt(\frac13\theta(x-x_0)+\frac{2}{3}\pi\rt),  x> 0.
\end{array}
\right.
\end{equation}
The stability potential corresponding to this sector is 
{\small \be 
{v}_c(x)=
\left\{
\begin{array}{c}
-\frac{10}{9}+22\,\wt{\chi}_3^2(x)-\frac{200}{3}\wt{\chi}_3^4(x)+\frac{448}{9}\wt{\chi}_3^6(x),\;x\leq 0,
\\ \\
-\frac{10}{9}+22\,\wt{\chi}_1^2(x)-\frac{200}{3}\wt{\chi}_1^4(x)+\frac{448}{9}\wt{\chi}_1^6(x), x> 0, 
\end{array}
\right.
\ee}where $\wt{\chi}_1(x)=\cos\lt(1/3 \arccos (\tanh(x-x_{0}))+2\pi/3\rt)$ and $\wt{\chi}_3(x)=\cos\lt(1/3 \arccos (\tanh(x+x_{0}))\rt)$.

\begin{figure}[ht]
\includegraphics[{height=3cm,width=7cm}]{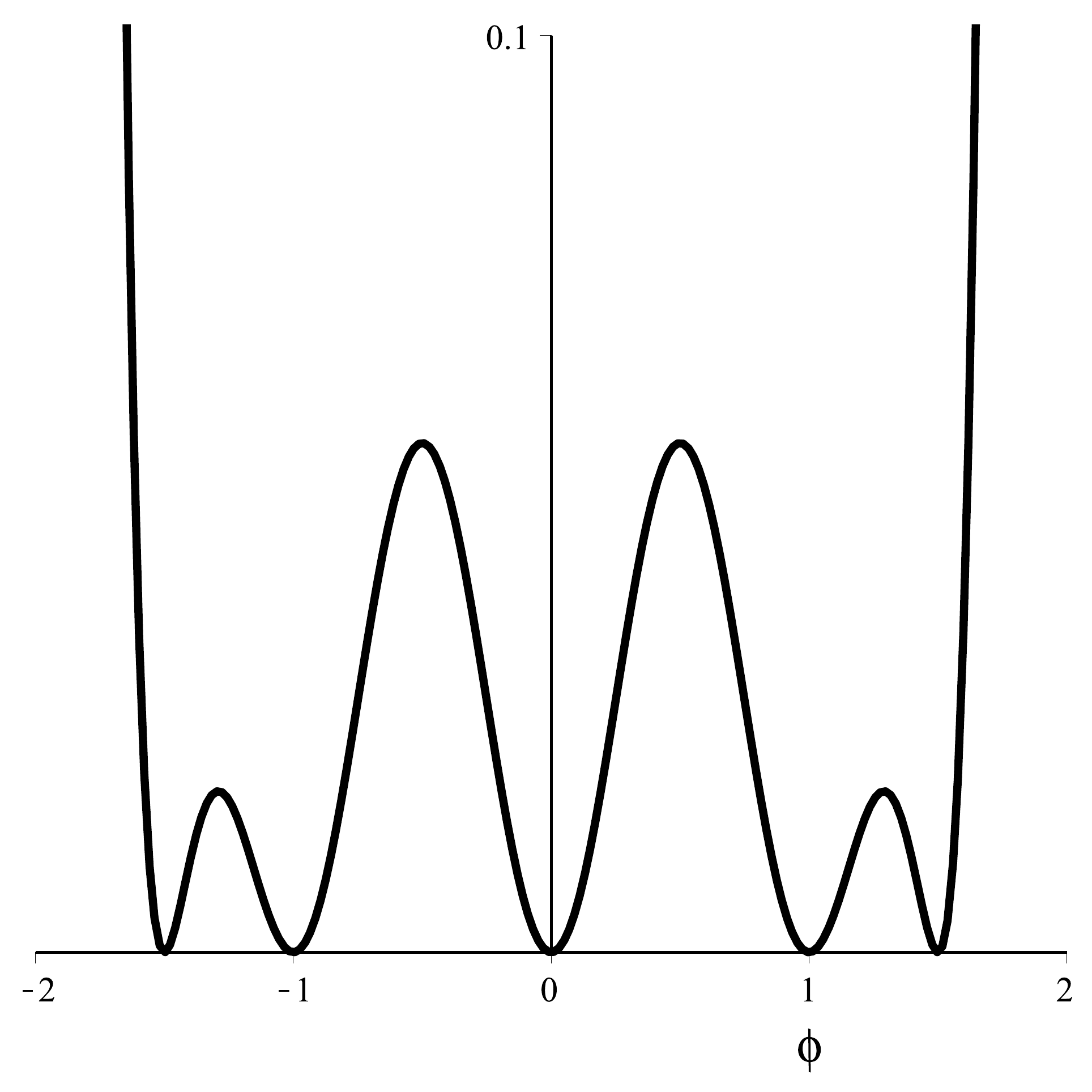}
\caption{The potential \eqref{po11} as a function of the field $\phi$, for $c=1/2$.}\label{fig9}
\end{figure}

For $c=1/2$, the potential \eqref{v81} is written as
\be
\label{po11}
V(\phi)=\frac1{18}\phi^2\left(1-|\phi|\right)^2\left(3-4\phi^2+4|\phi|\right)^2\,,
\ee
that has four topological sectors, as illustrated in Fig.~\ref{fig9}, and its static kink  solutions are
\bens
\phi_i(x)&=&-\dfrac{1}{2}+\chi_i(x)\,,\,\,\,\,i=(1,2)\,,\\
\phi_j(x)&=&\dfrac{1}{2}+\chi_{j-1}(x)\,,\,\,\,\,j=(3,4)\,,
\eens
with $\chi_k(x)$ given by Eqs.~\eqref{sol81}.
The energy densities and quantum potentials are the same given by  Eqs.~\eqref{rhop3} and \eqref{u8}, respectively. For all these four sectors, the energy densities and the stability potentials are as in the corresponding previous cases.

\subsubsection{Family 1}

Generalizing these results, we consider a family of non-deformed potentials with polynomial interactions,  firstly proposed in Ref.~\cite{Guilarte}. Let us consider
\be
\label{potfp}
{U}_{m}^{(1)}(\chi)=\frac{1}{2m^2}(1-\chi^2)^2{\bf U}_{m-1}^2(\chi),
\ee
where $m=1,2,3,4,...$  and ${\bf U}_{m-1}(\chi)$ represents the Chebyshev polynomials of second kind, given by
\be
\label{Chebyshev}
{\bf U}_{m-1}(\chi)=\frac{\sin(m\, \arccos(\chi))}{\sin(\arccos(\chi))}.
\ee
For each specific value of $m$, the potential \eqref{potfp} has $m$ topological sectors and $m+1$ minima given by
\be
\chi_{\min}^{(1)}=\cos\lt(\frac{k\pi}{m}\rt),
\ee
where $k=0,1,..,m$. The static solutions describing these sectors are
\be
\label{solfp}
\chi_{m,n}^{(1)}(x)=\cos\left(\frac1m\theta(x)+\frac{(m-n)}{m}\pi \right),
\ee
where  $n$ $(= 1,...,m)$ furnishes the kink solutions  connecting distinct topological sectors of ${U}_{m}^{(1)}(\chi)$. The energy densities of the solutions are given by
\be
\label{rhop}
{\epsilon}_{m,n}^{(1)}(x)=\frac{1}{m^2}\sech(x)^2\sin^2\left(\frac1m\theta(x)+\frac{(m-n)}{m}\pi \right).
\ee
Using Eqs.~\eqref{potfp} and \eqref{solfp}, we obtain the quantum potentials from Eq.~\eqref{Ut}, that is
\ben
\label{ufa}
{u}_{m,n}^{(1)}(x)&=&\frac{3}{m}\cot\lt(\frac1m\theta(x)+\frac{(m-n)}{m}\pi\rt)\sech(x)\tanh(x)
+1-\lt(2+\frac1{m^2}\rt)\sech^2(x).
\een

Applying the deformation function \eqref{def1} to the general potential \eqref{potfp}, we obtain de deformed model 
\be\label{vfa}
V_{m}^{(1)}(\phi)=\frac{1}{2m^2}\left(1-c^2+2c|\phi|-\phi^2\right)^2{\bf U}_{m-1}^2(c-|\phi|)\,.
\ee
Each value of $m$ gives a smooth function of $\phi$ when $c$ equals a minimum or maximum of ${U}_{m}^{(1)}(\chi)$. The expression above establishes  a set of polynomial potentials controlled by the parameters $m$ and $c$,  which provides new models having energy densities and quantum mechanical potentials like the ones obtained by ${U}_{m}^{(1)}(\chi)$, see Eqs.~\eqref{rhop} and \eqref{ufa}. The examples studied above are particular cases of this general family of potentials. 

\subsection{Hyperbolic models}

\subsubsection{Example 4}

\begin{figure}[ht]
\includegraphics[{height=3cm,width=7cm}]{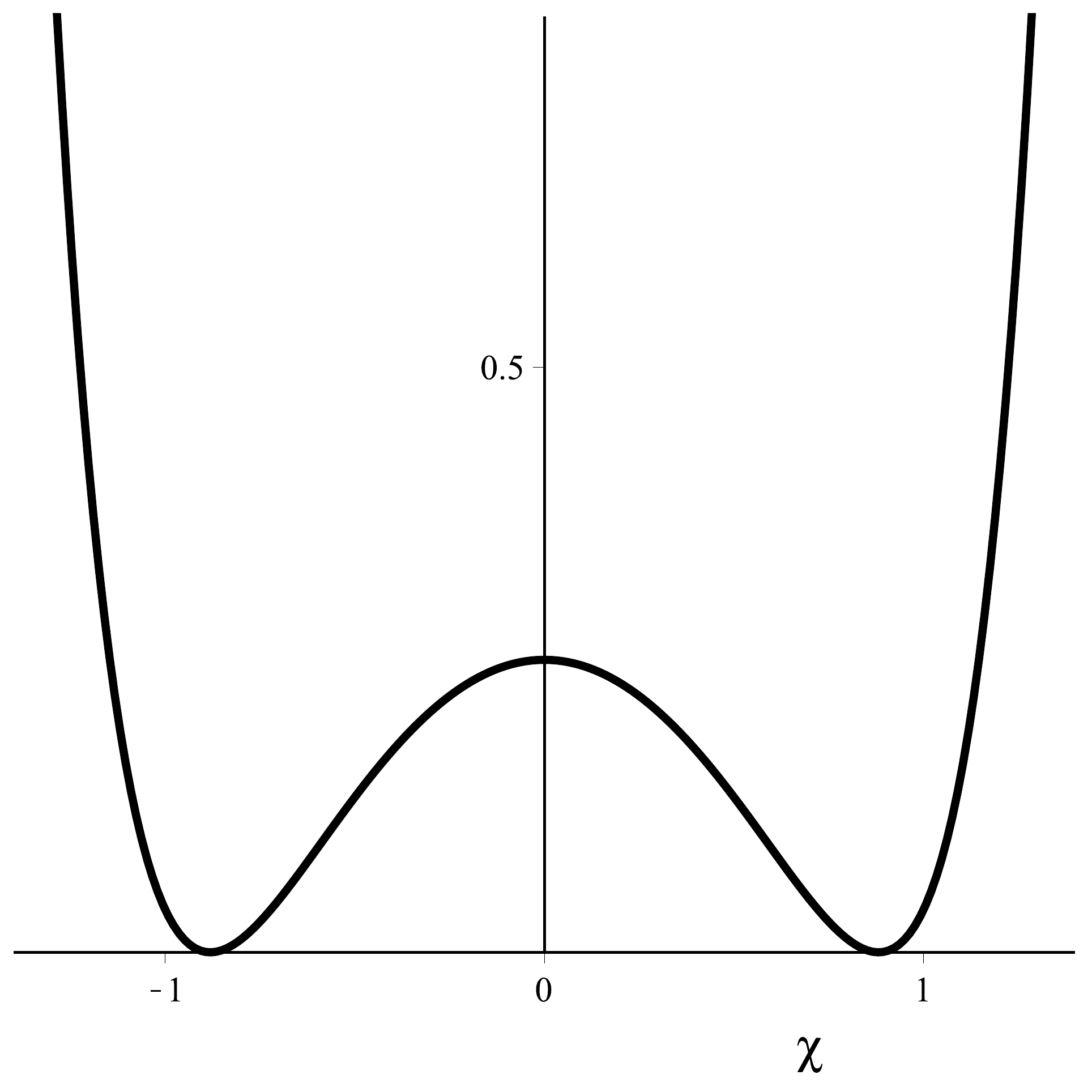}
\caption{The potential \eqref{vh1} as a function of the field $\chi$.}\label{fig10}
\end{figure}

\begin{figure}[ht]
\includegraphics[{height=3cm,width=7cm}]{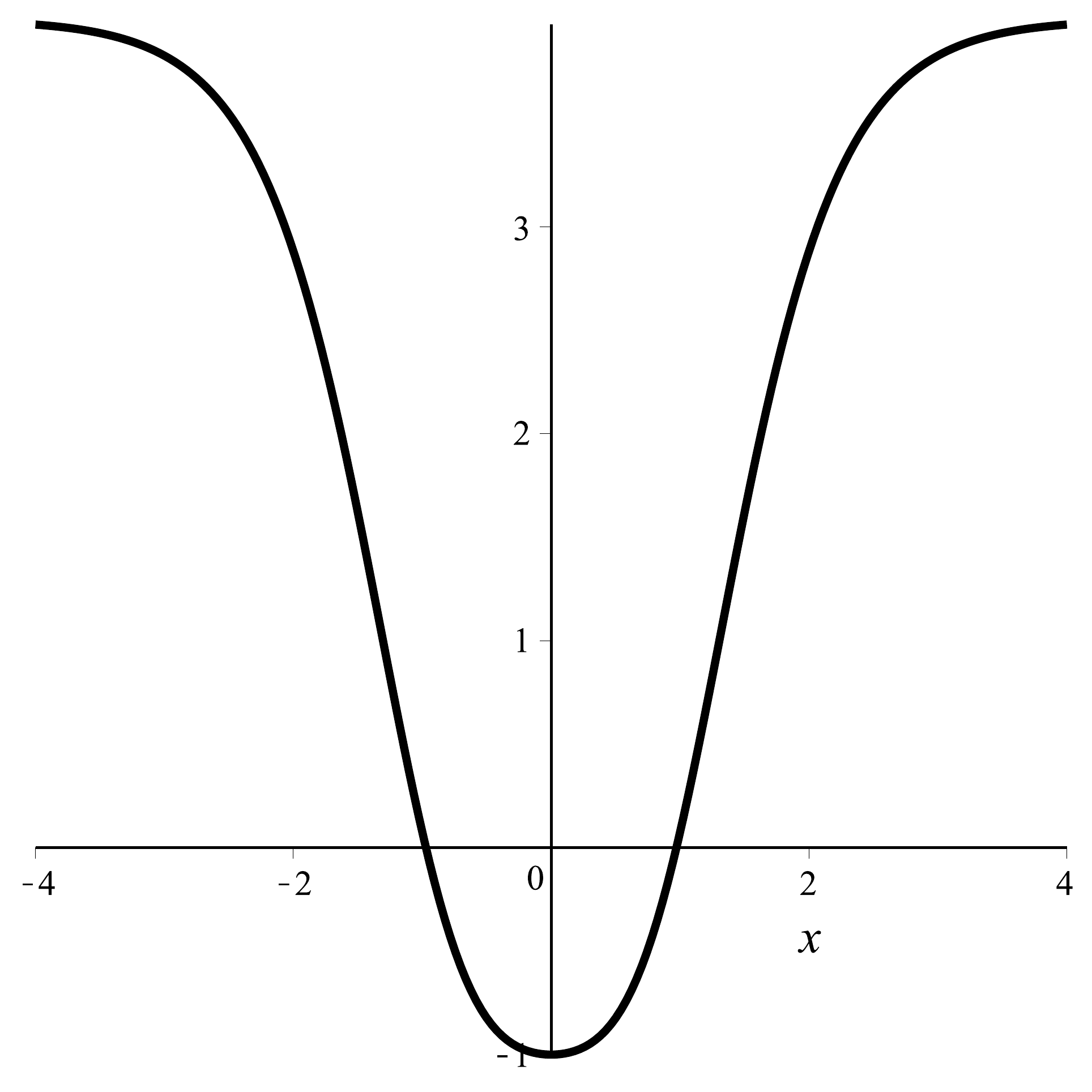}
\caption{The stability potential \eqref{Uh1}.}\label{figu10}
\end{figure}

Let us now start from potentials of the hyperbolic type. Take, for instance, the potential
\be\label{vh1}
{U}(\chi)=\dfrac{1}{4}\lt(1-\sinh^2 (\chi)\rt)^2\,,
\ee
which has the two minima $\chi_{\min}=\pm \arcsinh(1)$ and one local maximum at the origin, as illustrated in Fig.~\ref{fig10}. This model has one topological sector with kink solution
\be
\chi(x)=\arcsinh\lt(\frac{\tanh(x)}{\sqrt{2-\tanh^2(x)}}\rt).
\ee
Using some relations between inverse hyperbolic functions, it is possible to rewrite the solution above as
\be
\chi(x)={ \rm arctanh}\left(\dfrac{1}{\sqrt{2}}\tanh(x)\right)\,.
\ee
The energy density corresponding to this solution is
\be
{\epsilon}(x)=\frac{2\,\sech^4(x)}{\lt(2-\tanh^2(x)\rt)^2}.\\
\ee
This system admits the quantum potential
\ben\label{Uh1}
{u}(x)&=&1-2\,\sech^2(x)+\frac{3\,\tanh^4(x)}{2-\tanh^2(x)}
-\sech^2(x)\tanh^2(x)\frac{\lt(6+\tanh^2(x)\rt)}{\lt(2-\tanh^2(x)\rt)^2},
\een
which is depicted in Fig.~\ref{figu10}. 

\begin{figure}[ht]
\includegraphics[{height=3cm,width=7cm}]{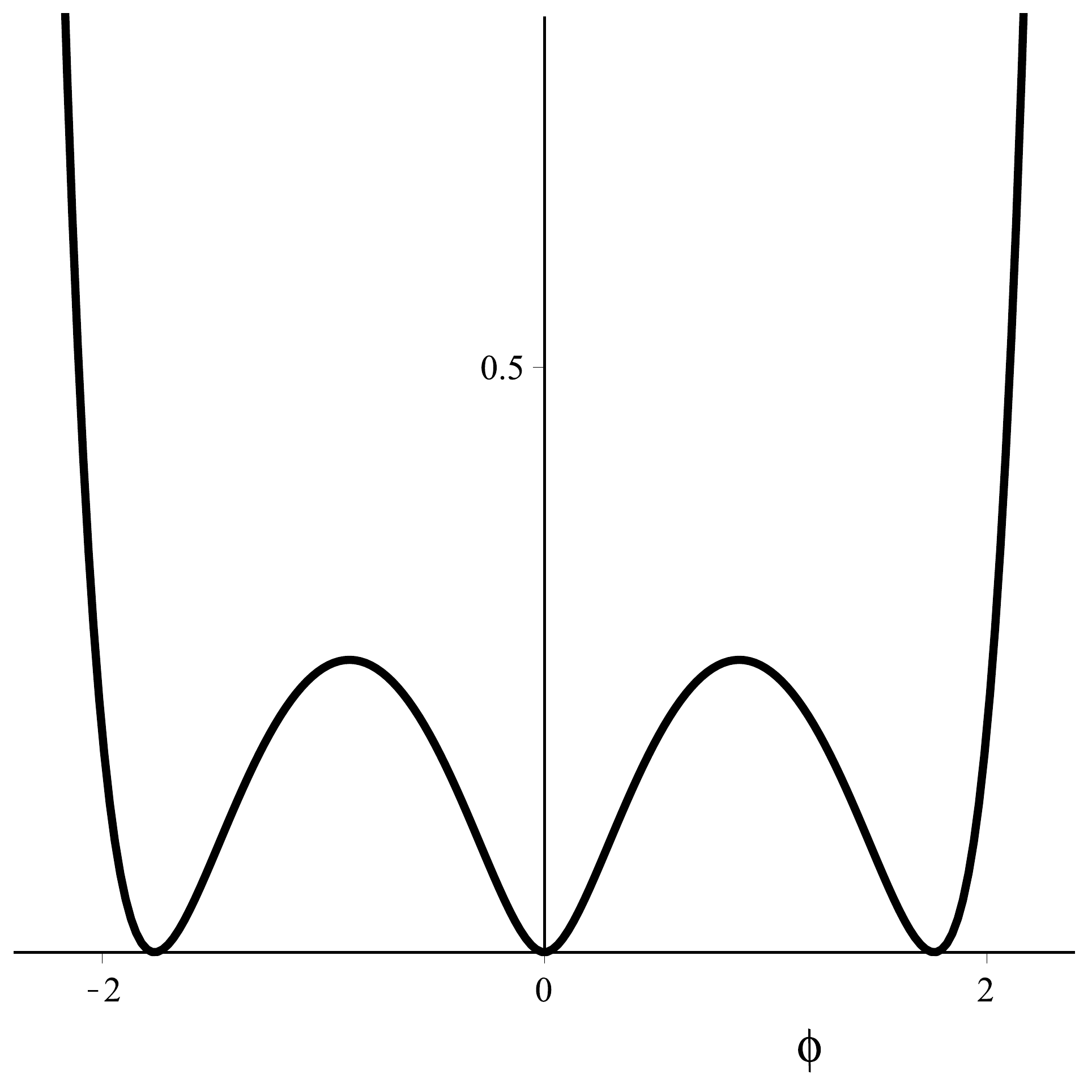}
\caption{The potential \eqref{vh1d} as a function of the field $\phi$, for $c= \arcsinh(1)$.}\label{fig11}
\end{figure}

 We can apply the deformation function \eqref{def1} to the potential \eqref{vh1},  then  we obtain the deformed potential 
\be
\label{vh1d}
V(\phi)=\dfrac{1}{4}\lt(1-\sinh^2 (c-|\phi|)\rt)^2\,,
\ee
which is a smooth function for $c= \arcsinh(1)$. When the parameter $c$ is equal to this value, two topological sectors appear as illustrated in Fig.~\ref{fig11}. The static kink  solutions are 
\bens
\phi_1(x)&=&-\arcsinh(1)+{ \rm arctanh}\left(\dfrac{1}{\sqrt{2}}\tanh(x)\right),\;\;\;\;\;\;\;\\
\phi_2(x)&=&\arcsinh(1)+{ \rm arctanh}\left(\dfrac{1}{\sqrt{2}}\tanh(x)\right). \;\;\;\;\;\;\;
\eens
It is possible to verify from Eq.~\eqref{U} that the quantum mechanical potential obtained for both sectors of this deformed model presents the same shape of ${u}(x)$ given by Eq.~\eqref{Uh1}.

\subsubsection{Example 5}

\begin{figure}[ht]
\includegraphics[{height=3cm,width=7cm}]{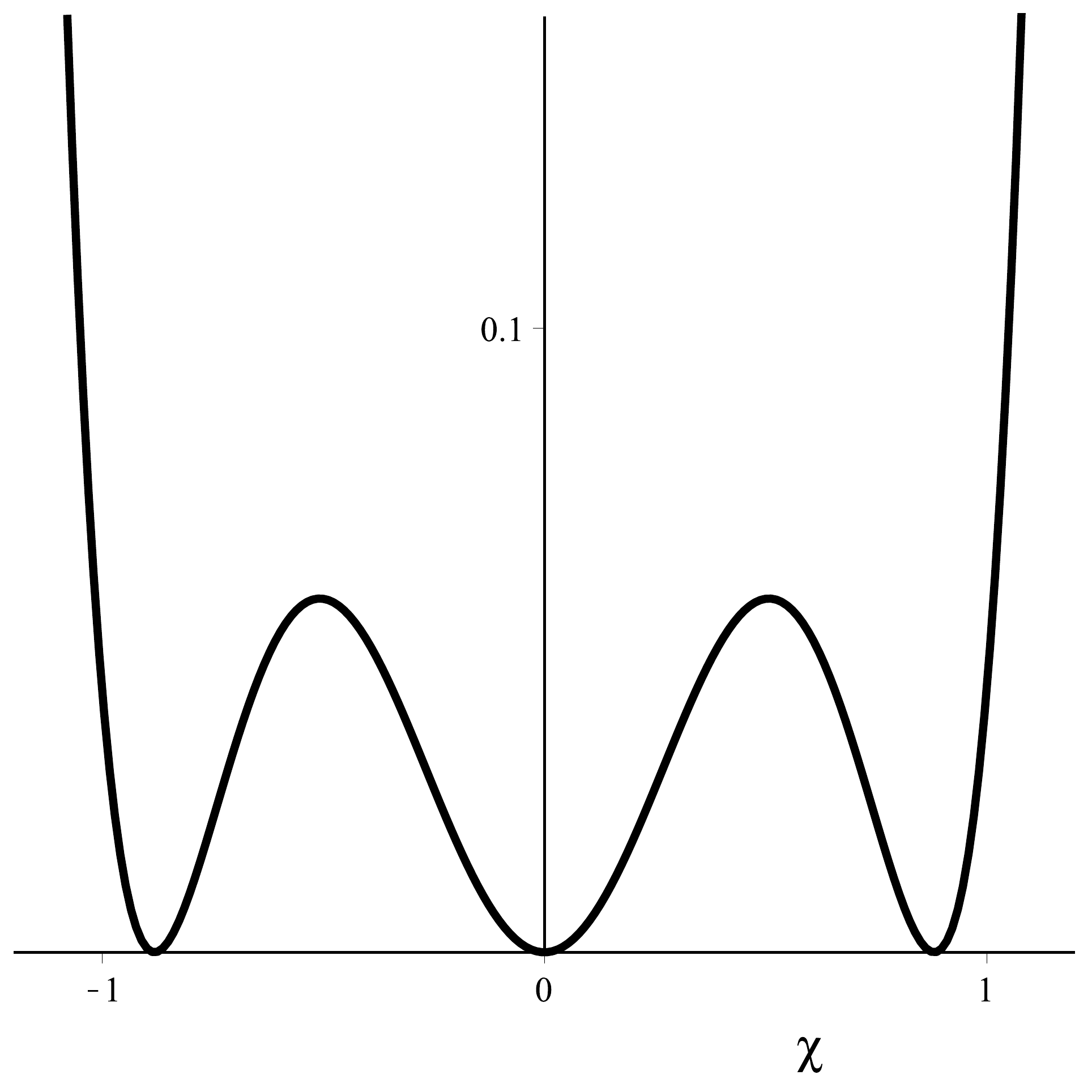}
\caption{The potential \eqref{vh2} as a function of the field $\chi$.}\label{fig12}
\end{figure}

\begin{figure}[ht]
\includegraphics[{height=3cm,width=7cm}]{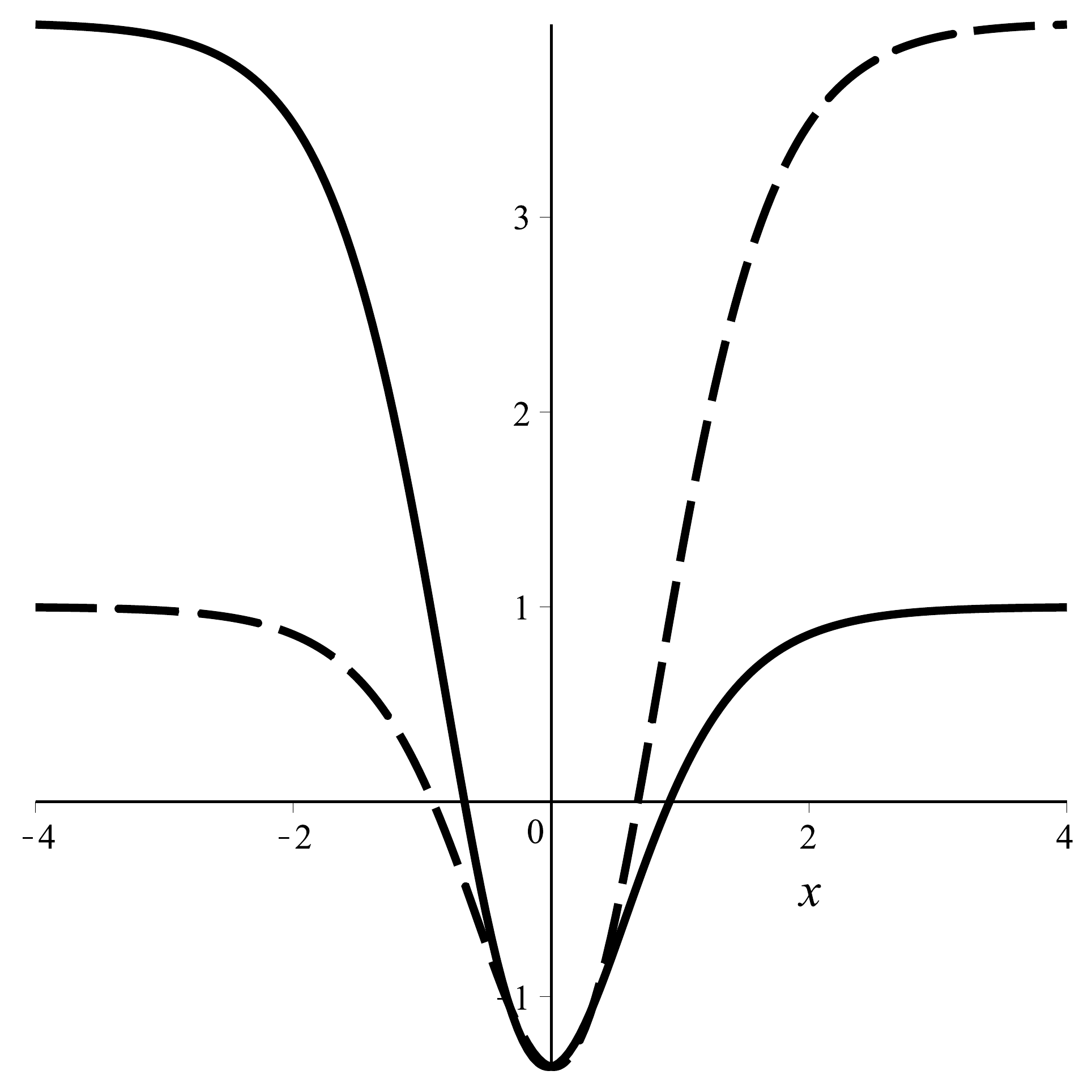}
\caption{The stability potentials  \eqref{Uh2a} (solid line) and \eqref{Uh2b} (dashed line).}\label{figu12}
\end{figure}

Here, we start with the potential 
\be
\label{vh2}
{U}(\chi)=\dfrac{1}{2}\tanh^{2}(\chi)(1-\sinh^2 (\chi))^2\,,
\ee
where the minima are $\chi_{\min}=0, \pm \arcsinh(1)$ and the maxima are approximately $\chi_{\max}= \pm 1/2$,  as illustrated in Fig.~\ref{fig12}. This model has two topological sectors with kink solutions given by
\bens
\label{solh2}
\chi_1(x)&=&-{\rm arctanh} \left(\dfrac{1}{\sqrt{2}}\sin\lt(\frac12\theta(x)\rt) \right)\,,\\
\chi_2(x)&=&{\rm arctanh} \left(\dfrac{1}{\sqrt{2}}\cos\left(\frac12\theta(x)\right)\right)\,,
\eens
where $\theta(x)$ is the same defined in the previous subsection. The corresponding energy densities are
\bens\label{eneh2}
{\epsilon}_1(x)=\frac{\cos^2\lt(\frac12\theta(x)\rt)}{2\lt(2-\sin^2\lt(\frac12\theta(x)\rt)\rt)^2}\sech^2(x),\\
{\epsilon}_2(x)=\frac{\sin^2\lt(\frac12\theta(x)\rt)}{2\lt(2-\cos^2\lt(\frac12\theta(x)\rt)\rt)^2}\sech^2(x).
\eens
The associated stability potentials are given by, respectively
\bens \label{Uh2}
\slabel{Uh2a}
{u}_1(x)&=& 1-2\,\sech^2(x)- \frac{\sin^2\lt(\frac12\theta(x)\rt)\tan\lt(\frac12\theta(x)\rt)}{2-\sin^2\lt(\frac12\theta(x)\rt)}
\times\frac32\sech(x)\tanh(x) \nonumber\\
&&-\frac{\lt(6+\sin^2\lt(\frac12\theta(x)\rt)\rt)}{\lt(2-\sin^2\lt(\frac12\theta(x)\rt)\rt)^2}
\times\frac14\sin^2\lt(\frac12\theta(x)\rt)\sech^2(x),\\
\slabel{Uh2b}
{u}_2(x)&=& 1-2\,\sech^2(x)+ \frac{\cos^2\lt(\frac12\theta(x)\rt)\cot\lt(\frac12\theta(x)\rt)}{2-\cos^2\lt(\frac12\theta(x)\rt)}
\times\frac32\sech(x)\tanh(x) \nonumber\\
&&-\frac{\lt(6+\cos^2\lt(\frac12\theta(x)\rt)\rt)}{\lt(2-\cos^2\lt(\frac12\theta(x)\rt)\rt)^2}
\times\frac14\cos^2\lt(\frac12\theta(x)\rt)\sech^2(x),
\eens
which are depicted in Fig.~\ref{figu12}.

By considering the deformation  \eqref{def1} in the potential \eqref{vh2}, we get 
\be\label{vhd2}
V(\phi)=\dfrac{1}{2}\tanh^{2}\lt(c-|\phi|\rt)\lt(1-\sinh^2 \lt(c-|\phi|\rt)\rt)^2.
\ee 
This potential is a smooth function for $c=\arcsinh(1)$, and $c \approx 1/2$, as we will consider below.

\begin{figure}[ht]
\includegraphics[{height=3cm,width=7cm}]{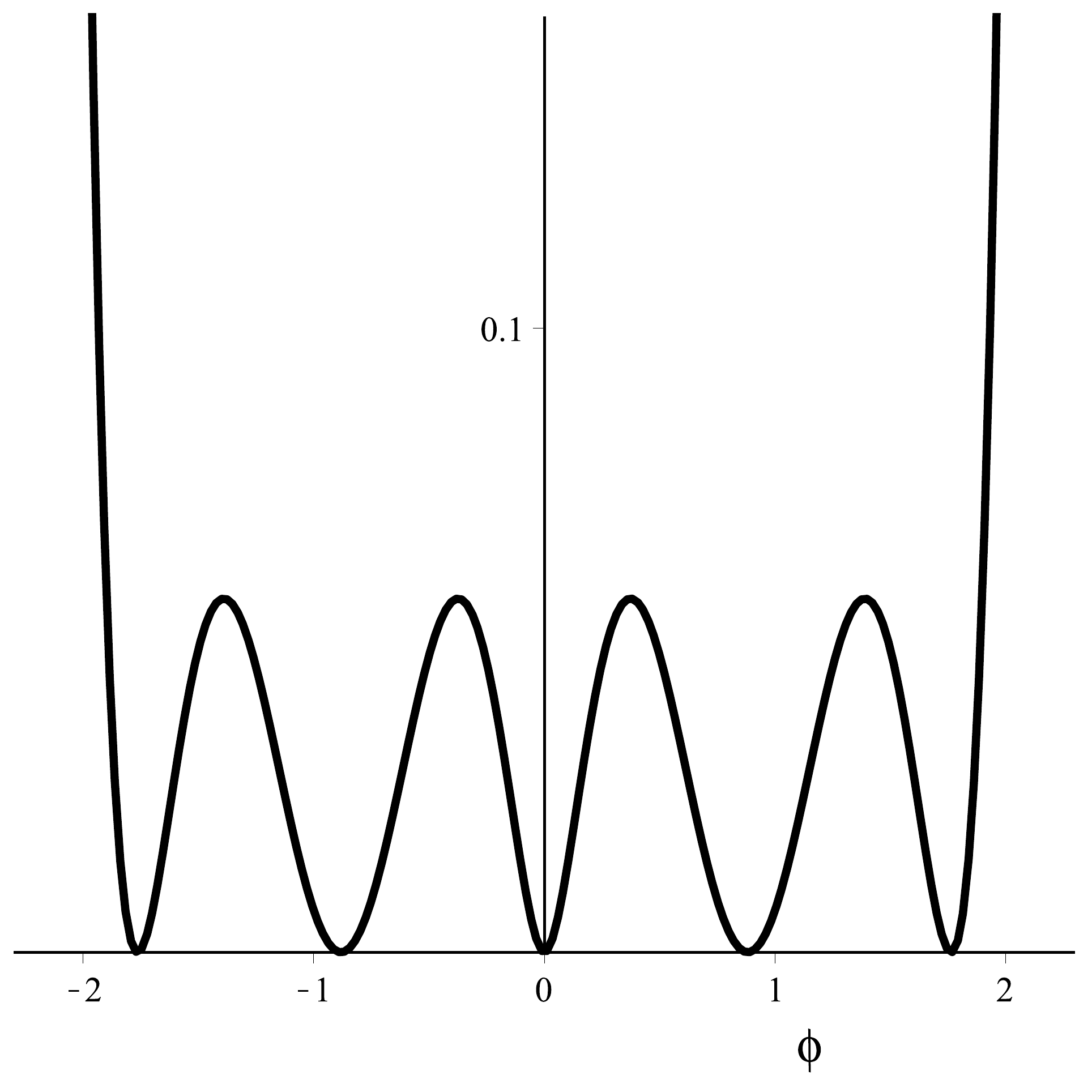}
\caption{The potential \eqref{vhd2} as a function of the field $\phi$, for $c=\arcsinh(1)$.}\label{fig13}
\end{figure}

For $c= \arcsinh(1)$, four topological sectors appear as illustrated in Fig.~\ref{fig13}. These sectors from left to right, respectively, present the static kink  solutions  
\bens
\phi_i(x)&=&-\arcsinh(1)+\chi_i(x)\,,\,\,\,\,\, i=(1,2)\,,\\
\phi_j(x)&=&\arcsinh(1)+\chi_{j-2}(x)\,\,,j=(3,4),
\eens
with $\chi_k(x)$ given by eqs. \eqref{solh2}. The energy densities 
are also given by Eq.~\eqref{eneh2}, and the stability potentials are similar to those shown in Eq.~\eqref{Uh2}, appropriately.

\begin{figure}[ht]
\includegraphics[{height=3cm,width=7cm}]{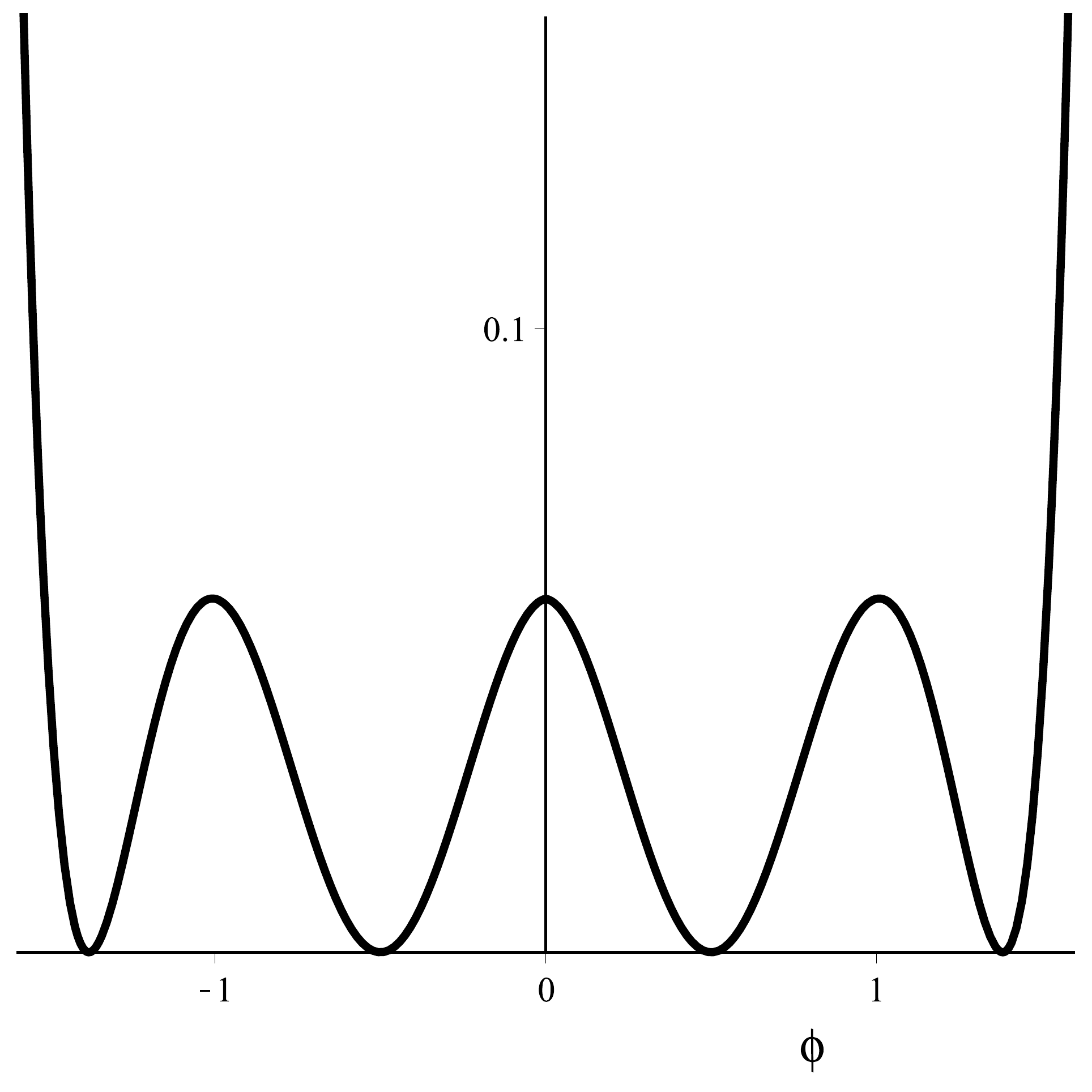}
\caption{The potential \eqref{vhd2} as a function of the field $\phi$, for $c= 1/2$.}\label{fig14}
\end{figure}

\begin{figure}[ht]
\includegraphics[{height=3cm,width=7cm}]{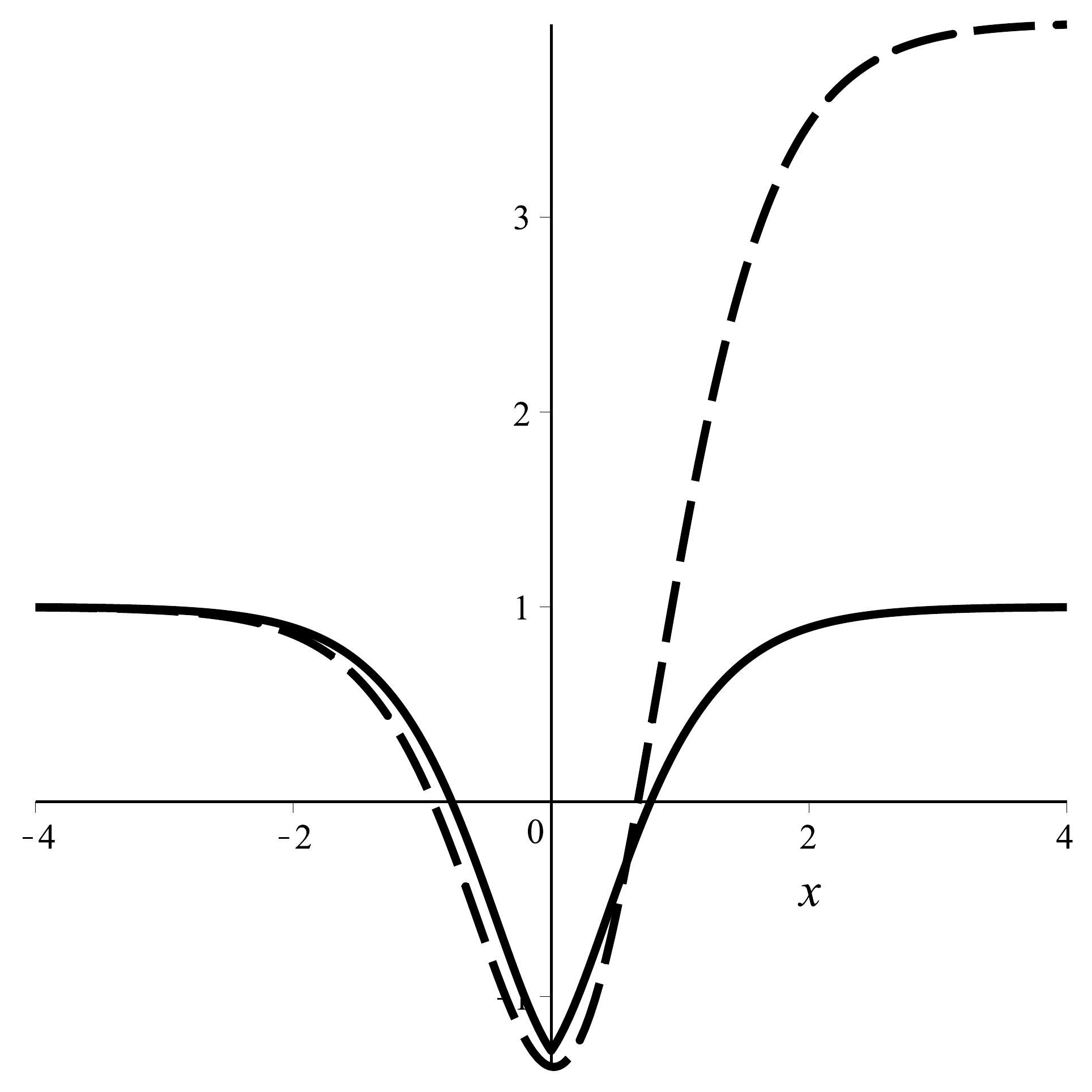}
\caption{The stability potentials \eqref{Uhc1} (solid line) and \eqref{Uh2b} (dashed line).}\label{figu14}
\end{figure}

For $c\approx 1/2$, the model \eqref{vhd2} exhibits three topological sectors as displayed in Fig.~\ref{fig14}. The two lateral sectors have static kink  solutions given by  
\bens
\phi_1(x)&=&-\dfrac{1}{2}+\chi_1(x)\,,\\
\phi_3(x)&=&\dfrac{1}{2}+\chi_2(x)\,,
\eens
with $\chi_k(x)$ given by eqs. \eqref{solh2}, and  the central sector has static solution 
\begin{equation}
{\phi}_c(x)=
\left\{
\begin{array}{c}
-\frac{1}{2}+ {\rm arctanh} \left(\frac{1}{\sqrt{2}}\cos \left(\frac{1}{2}\theta(x-x_0)\right)\right), \; x\leq 0\,,\\
\,\,\\
 \;\; \frac{1}{2}- {\rm arctanh} \left(\frac{1}{\sqrt{2}}\sin \left(\frac{1}{2}\theta(x+x_0) \right)\right), \;\;  x> 0\,,
\end{array}
\right.
\end{equation}
where $x_0\approx {\rm arctanh}\lt(\cos\lt(2 \arcsin (\sqrt{2} \tanh(1/2))\rt)\rt)$.
The energy densities are given in terms of Eq.~\eqref{eneh2}, for the lateral sectors $\rho_1(x)={\epsilon}_1(x)$ and $\rho_3(x)={\epsilon}_2(x)$, while for the central one
\ben \label{enehc1}
\rho_c(x)= 
\left\{
\begin{array}{c}
{\epsilon}_{2}(x-x_0), \;\;\; x\leq 0\,,\\
\,\,\\
{\epsilon}_{1}(x+x_0),  \;\;\; x> 0\,.
\end{array}
\right.
\een
As a consequence, the stability potentials for the two lateral sectors are analogous to \eqref{Uh2}, that is, $v_{1}(x)={u}_{1}(x)$ and
$v_{3}(x)={u}_{2}(x)$. And for the central sector we have
\ben \label{Uhc1}
v_c(x)= 
\left\{
\begin{array}{c}
{u}_{2}(x-x_0), \;\;\; x\leq 0\,,\\
\,\,\\
{u}_{1}(x+x_0),  \;\;\; x> 0\,,
\end{array}
\right.
\een
which is displayed in Fig.~\ref{figu14} together with the potential \eqref{Uh2b}.

\subsubsection{Example 6}

\begin{figure}[ht]
\includegraphics[{height=3cm,width=7cm}]{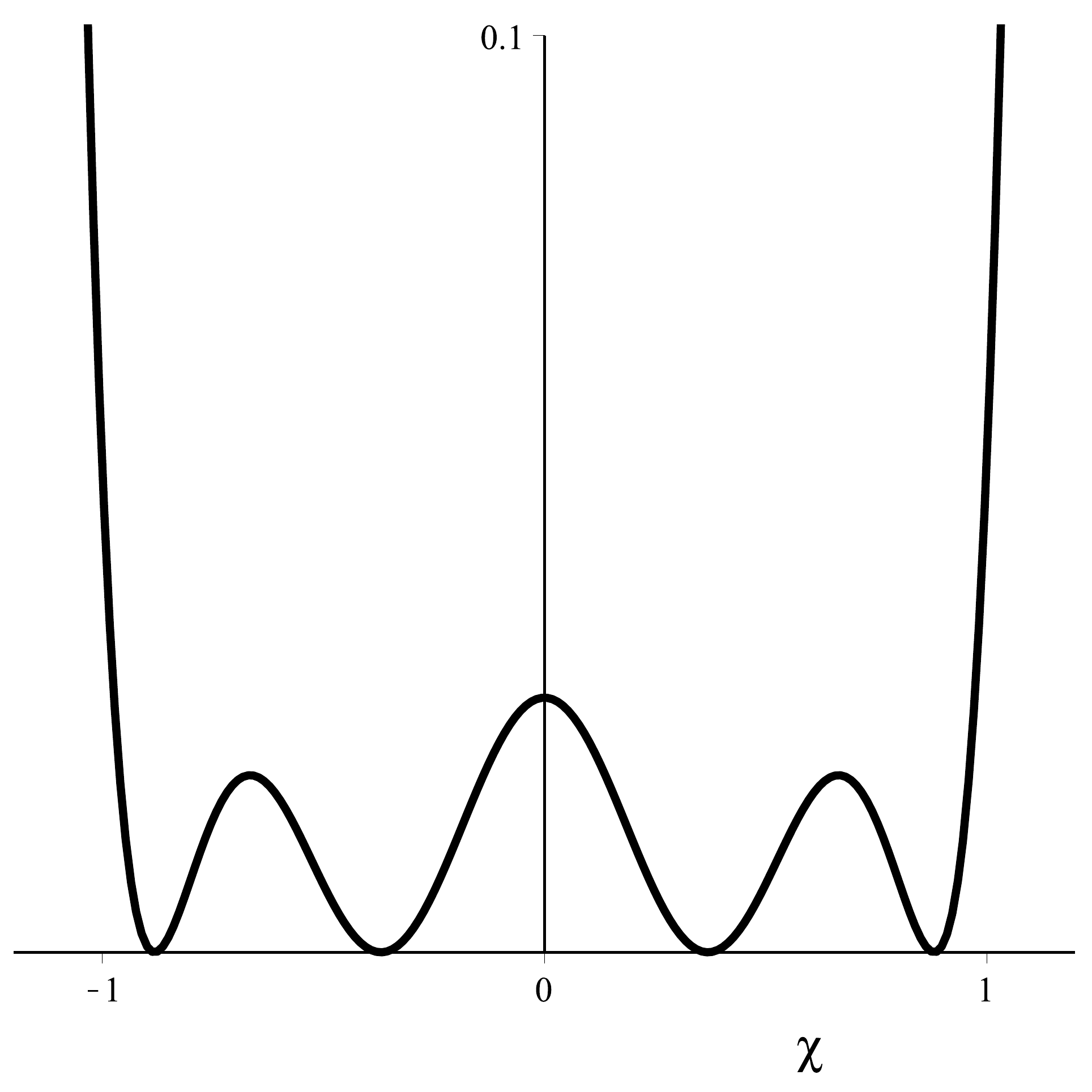}
\caption{The potential \eqref{vh3} as a function of the field $\chi$.}\label{fig15}
\end{figure}

\begin{figure}[ht]
\includegraphics[{height=3cm,width=7cm}]{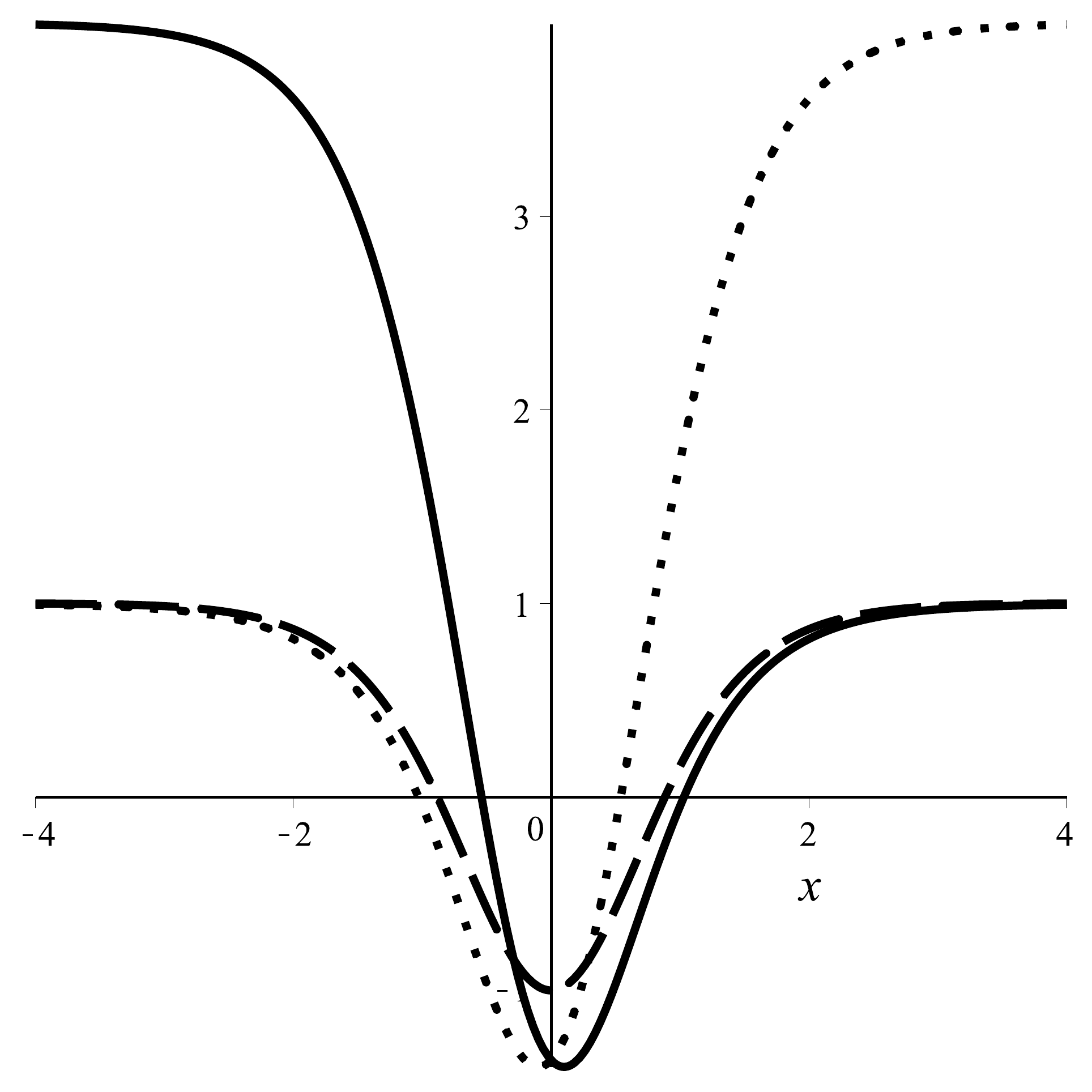}
\caption{The stability potential \eqref{Uh3}, for $n=1$ (solid line), $n=2$ (dashed line), and $n=3$ (dotted line).}\label{figu15}
\end{figure}

Let us consider one more example of hyperbolic potential
\be\label{vh3}
{U}(\chi)=\dfrac{1}{36} \sech^{4}(\chi)\lt(1-7\sinh^{2}(\chi)\rt)^2\lt(1-\sinh^{2}(\chi)\rt)^2, 
\ee
which has maxima at $\chi=0$ and $\chi\approx \pm 2/3$,  and 
minima at $\chi= \pm \arcsinh(1)$ and $\chi=\pm \arcsinh(1/\sqrt{7})$. This model has three topological sectors, as illustrated in Fig.~\ref{fig15}, with their respective kink solutions given by
\be\label{solh3}
\chi_n(x)={\rm arctanh} \left(\dfrac{1}{\sqrt{2}}\cos\left(\frac13\theta(x)+\frac{(3-n)}3\pi\right)\right)\,,
\ee
where $n=1,2,3$, whose energy densities are given by
\be
\label{enh1}
{\epsilon}_{n}(x)=\dfrac{2\sin^2\left(\dfrac13\theta(x)+\dfrac{(3-n)}{3}\pi\right)\sech^2(x)}{9\left(2-\cos^2\left(\dfrac13\theta(x)+\dfrac{(3-n)}{3}\pi\right)\right)^2}\,.
\ee
Also, the corresponding stability potentials are given by
\ben \label{Uh3}
{u}_n(x)&=&1-2\sech^2(x) + \cot\left(\frac13\theta(x)+\frac{(3-n)}{3}\pi\right)\times
\frac{\cos^2\left(\frac13\theta(x)+\frac{(3-n)}{3}\pi\right)}{2-\cos^2\left(\frac13\theta(x)+\frac{(3-n)}{3}\pi\right)} \sech(x)\tanh(x) \nonumber \\ & & 
-\frac19 \sech^2(x)\cos^2\left(\frac13\theta(x)+\frac{(3-n)}{3}\pi\right) \times 
\frac{6+\cos^2\left(\frac13\theta(x)+\frac{(3-n)}{3}\pi\right)}{\left(2-\cos^2\left(\frac13\theta(x)+\frac{(3-n)}{3}\pi\right)\right)^2}, 
\een
which is plotted in Fig.~\ref{figu15} for $n=1,2,3$.

Assuming the deformation function \eqref{def1} applied to the potential \eqref{vh3},  we obtain the deformed potential given by
\ben\label{vhd3}
V(\phi)&=&\dfrac{1}{36} \sech^{4}(c-|\phi|)\lt(1-7\sinh^{2}(c-|\phi|)\rt)^2
\lt(1-\sinh^{2}(c-|\phi|)\rt)^2,
\een
which is a smooth function for $c=\arcsinh(1)$, $2/3$, and $\arcsinh(1/\sqrt{7})$, as we will consider below.

\begin{figure}[ht]
\includegraphics[{height=3cm,width=7cm}]{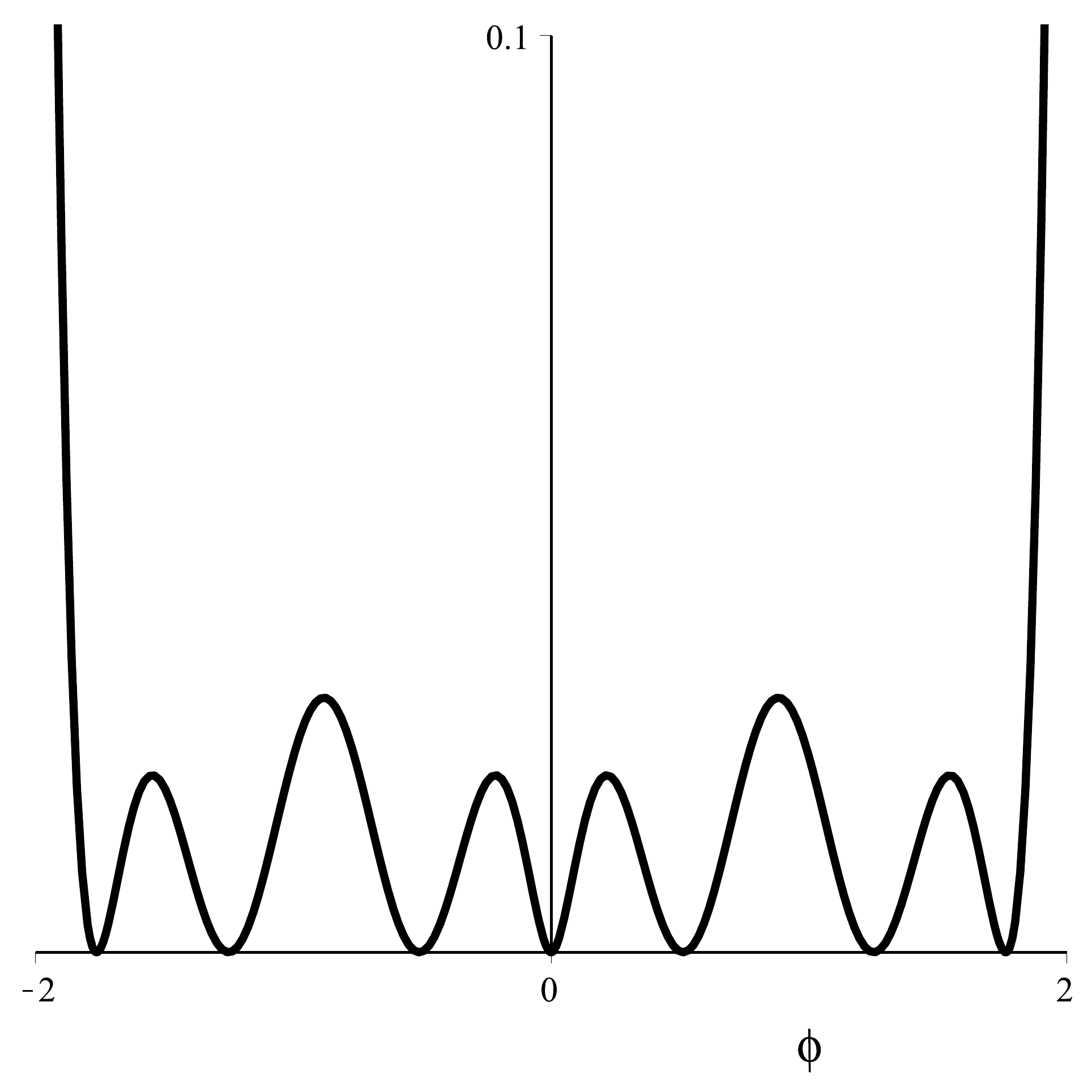}
\caption{The potential \eqref{vhd3} as a function of the field $\phi$, for $c= \arcsinh(1)$.}\label{fig16}
\end{figure}

For $c=\arcsinh(1)$,  six topological sectors appear as illustrated in Fig.~\ref{fig16}. These sectors  present the following static kink  solutions  
\bens
\phi_i(x)&=&-\arcsinh(1)+\chi_i(x)\,,\,\,\, i=(1,2,3)\,,\\
\phi_j(x)&=&\arcsinh(1)+\chi_{j-3}(x)\,,j=(4,5,6),
\eens
with $\chi_k(x)$ given by Eqs.~\eqref{solh3}. The associated energy densities are given in Eq.~\eqref{enh1}, and the stability potentials are equivalent to those seen in Eq.~\eqref{Uh3}.

\begin{figure}[ht]
\includegraphics[{height=3cm,width=7cm}]{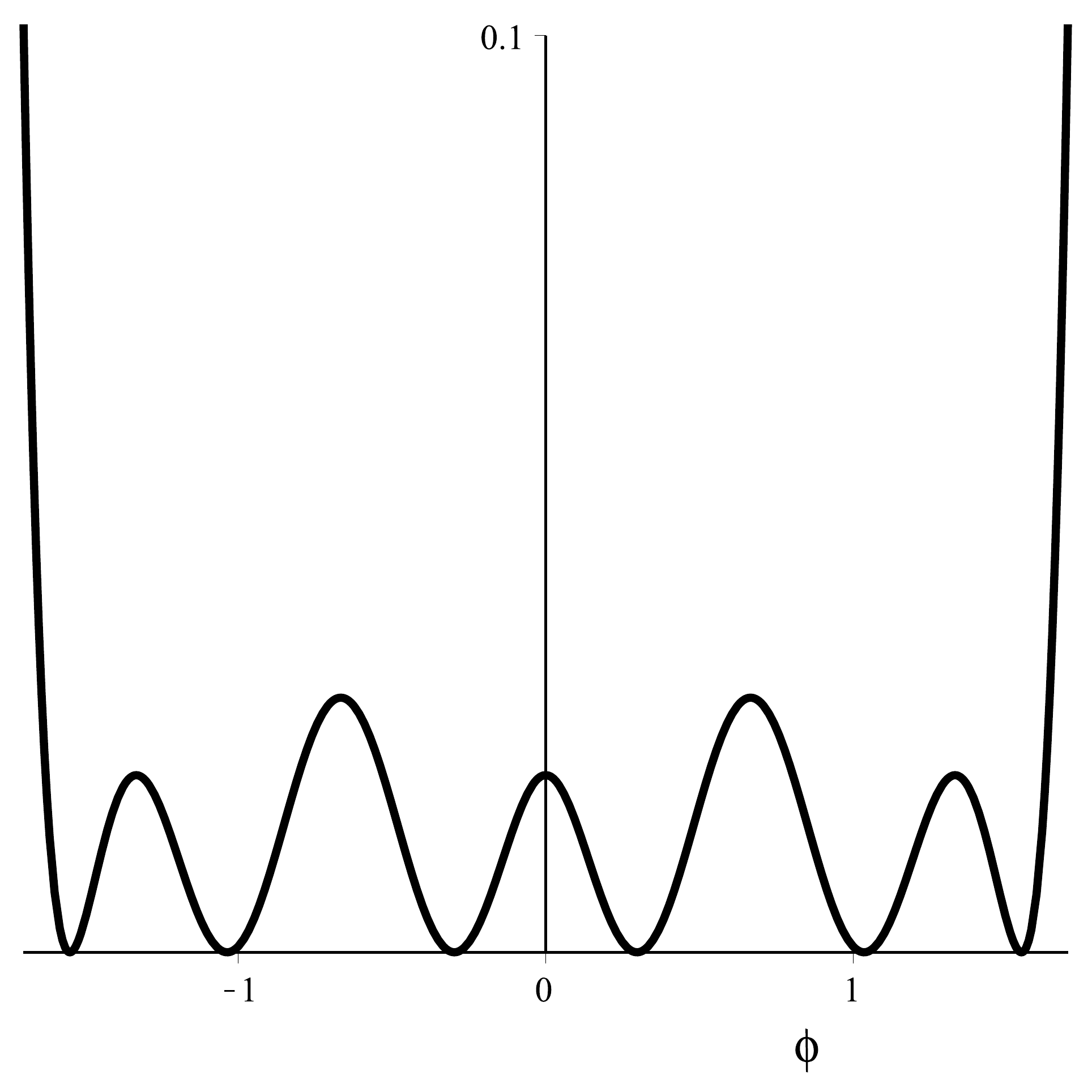}
\caption{The potential \eqref{vhd3} as a function of the field $\phi$, for $c= 2/3$.}\label{fig17}
\end{figure}

For $c= 2/3$, the model \eqref{vhd3} presents five topological sectors as displayed in Fig.~\ref{fig17}. The  static kink  solutions without the one related to the central sector  are
\bens
\phi_i(x)&=&-\dfrac{2}{3}+\chi_i(x)\,,\,\,\,i=(1,2)\,,\\
\phi_j(x)&=&\dfrac{2}{3}+\chi_{j-2}(x)\,,j=(4,5)\,,
\eens
with $\chi_k(x)$ given by Eqs.~\eqref{solh3},
which provide  the energy densities \eqref{enh1} and  the stability potentials \eqref{Uh3}; however, for the central topological sector, we get the static kink solution 
\begin{equation}
{\phi}_c(x)=
\left\{
\begin{array}{c}
-\frac{2}{3}+{ \rm arctanh }\left(\frac{1}{\sqrt{2}}\cos \left(\frac{1}{3}{\theta}_1(x)\right)\right),\;\;\;\;\;\; x\leq 0\,,\\
\\
\frac{2}{3}+ {\rm arctanh} \left(\frac{1}{\sqrt{2}}\cos \left(\frac{1}{3}{\theta}_2(x)+\frac{2\pi}{3}\right)\right), \; x> 0\,,
\end{array}
\right.
\end{equation}
where  ${\theta}_1(x)=\theta(x+x_0) =\arccos(\tanh(x+x_0))$, ${\theta}_2(x)=\theta(x-x_0)=\arccos (\tanh(x-x_0))$, and $x_0={\rm arctanh}\lt(\cos\lt(3 \arccos(\sqrt{2}\tanh(2/3))\rt)\rt)$. The energy density has the form

\ben \label{eneh3}
\rho_c(x)= 
\left\{
\begin{array}{c}
{\epsilon}_{3}(x+x_0), \;\;\; x\leq 0\,,\\
\,\,\\
{\epsilon}_{1}(x-x_0),  \;\;\; x> 0\,,
\end{array}
\right.
\een
where ${\epsilon}_3(x+x_0)$ and ${\epsilon}_1(x-x_0)$ are given by Eq.~\eqref{enh1} replacing $x\rightarrow x+x_0$ and $x\rightarrow x-x_0$, respectively. And the corresponding stability potential is
\ben \label{Uhc3}
v_c(x)= 
\left\{
\begin{array}{c}
{u}_{3}(x+x_0), \;\;\; x\leq 0\,,\\
\,\,\\
{u}_{1}(x-x_0),  \;\;\; x> 0\,,
\end{array}
\right.
\een
where ${u}_3(x+x_0)$ and ${u}_1(x-x_0)$ are given by Eq.~\eqref{Uh3}.

\begin{figure}[ht]
\includegraphics[{height=3cm,width=7cm}]{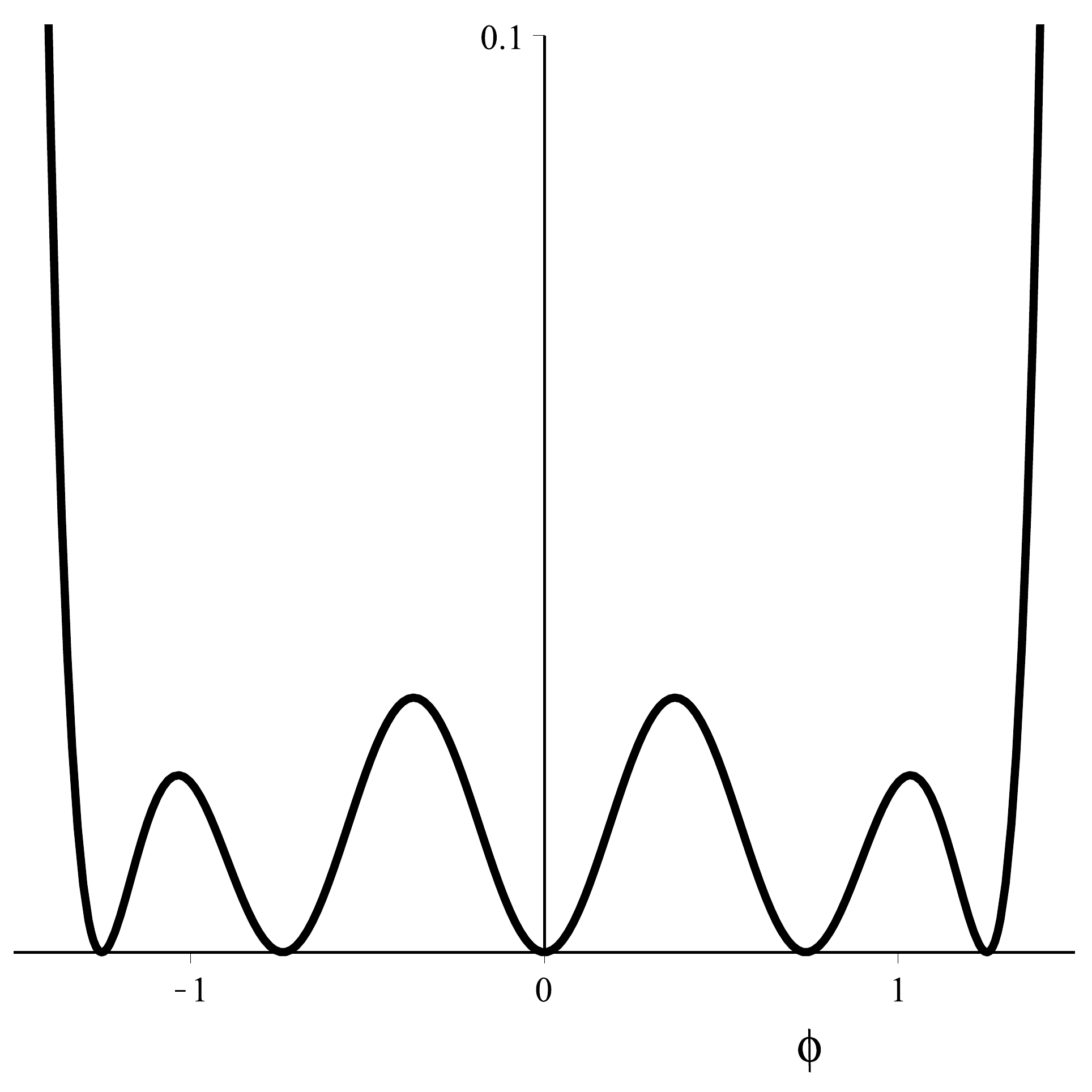}
\caption{The potential \eqref{vhd3} as a function of the field $\phi$, for $c=\arcsinh(1/\sqrt{7})$.}\label{fig18}
\end{figure}

For $c=\arcsinh(1/\sqrt{7})$,  the model \eqref{vhd3} presents four topological sectors as displayed in Fig.~\ref{fig18}. The  static kink  solutions are
\bens
\phi_i(x)&=&-\arcsinh(1/\sqrt{7})+\chi_i(x),\,\,\,i=(1,2),\;\\
\phi_j(x)&=&\arcsinh(1/\sqrt{7})+\chi_{j-2}(x),j=(3,4),\; 
\eens
with $\chi_k(x)$ given by Eqs.~\eqref{solh3}, which furnish the energy densities \eqref{enh1}, and the  quantum potentials \eqref{Uh3}.

\subsubsection{Family 2}

We can generalize the procedure and consider a family of potentials with hyperbolic interactions firstly studied in Ref.~\cite{Lima}. We take 
\be
\label{poth}
{U}_{m}^{(2)}(\chi)=\frac{1}{4m^2}\left(1-\sinh^2(\chi)\right)^2 {\bf U}_{m-1}^2\left(\sqrt{2}\,\tanh(\chi)\right),
\ee
where $m=1,2,3,...$  and ${\bf U}_{m-1}\left(\sqrt{2}\,\tanh(\chi)\right)$ represents Chebyshev polynomials of second kind. Each particular value of $m$ leads to $m+1$ minima that can be written as
\be
\chi_{\min}^{(2)}= {\rm arctanh}\left(\frac{1}{\sqrt{2}}\cos\left(\frac{k\pi}{m}\right)\right),
\ee
where $k=0,1,..,m$. The static kink solutions describing distinct topological sectors of ${U}_{m}^{(2)}(\chi)$ are
\be
\label{solh}
\chi_{m,n}^{(2)}(x)= {\rm arctanh} \left\{\frac{1}{\sqrt{2}}\cos\left(\frac1m\theta(x)+\frac{(m-n)}{m}\pi\right)\right\},
\ee
where  $n = 1,...,m$. The energy densities associated to these solutions are 
\be
\label{rhoph}
{\epsilon}_{m,n}^{(2)}(x)=\frac{2}{m^2}\frac{\sin^2\left(\frac1m\theta(x)+\frac{(m-n)}{m}\pi \right)\sech^2(x)}{\left(2-\cos^2\left(\frac1m\theta(x)+\frac{(m-n)}{m}\pi \right)\right)^2}\,.
\ee
After replacing Eqs.~\eqref{poth} and \eqref{solh} into Eq.~\eqref{Ut}, we obtain the following stability potentials
{\small
\ben \label{ufah}
{u}_{m,n}^{(2)}(x)&=&1-2\sech^2(x) + \frac3m  \cot\left(\frac1m\theta(x)+\frac{(m-n)}{m}\pi\right)\times 
 \frac{\cos^2\left(\frac1m\theta(x)+\frac{(m-n)}{m}\pi\right)}{2-\cos^2\left(\frac1m\theta(x)+\frac{(m-n)}{m}\pi\right)} \sech(x)\tanh(x) \nonumber \\ & & 
-\frac1{m^2} \sech^2(x)\cos^2\left(\frac1m\theta(x)+\frac{(m-n)}{m}\pi\right) \times
\frac{6+\cos^2\left(\frac1m\theta(x)+\frac{(m-n)}{m}\pi\right)}{\left(2-\cos^2\left(\frac1m\theta(x)+\frac{(m-n)}{m}\pi\right)\right)^2}. 
\een}

Applying the deformation function \eqref{def1} to the generic potential \eqref{poth}, we obtain de deformed model 
\ben\label{vfah}
V_{m}^{(2)}(\phi)&=&\frac{1}{4m^2}\left(1-\sinh^2(c-|\phi|)\right)^2 {\bf U}_{m-1}^2\left(\sqrt{2}\,\tanh(c-|\phi|)\right).
\een
We notice that for each value of $m$, it is a smooth function of $\phi$ when $c$ is equal to a minimum or maximum of ${U}_{m}^{(2)}(\chi)$. The expression above establishes  a set of hyperbolic potentials controlled by the parameters $m$ and $c$,  which provides new models presenting energy densities and stability potentials like the ones obtained by ${U}_{m}^{(2)}(\chi)$, see Eqs.~\eqref{rhoph} and \eqref{ufah}. The three examples studied above are particular cases of this general family of potentials.

\section{Comments and conclusions} \label{sec4}

In the current work we investigated standard field theories of a single real scalar field in (1+1)-dimensions,
and explored several aspects related to their stability potentials. The proposal presented here is inspired by the result recently obtained in \cite{BEN2}, which showed that it is possible to use the reconstruction procedure and start from a quantum mechanical potential and end up with two distinct field theory models, thus showing that the process of reconstruction of field theory from supersymmetric quantum mechanics is not unique. Also, it is motived by another recent result \cite{BL}, which unveiled the possibility of using the deformation procedure to write distinct field theory models, having the very same stability potential.   

Our goal here was to find distinct models presenting the same stability potentials and for this, we considered the deformation procedure  where the connection between these models is determined by the deformation function \cite{DP}. Then a relation is given establishing a connection that ensure the construction of distinct systems having similar stability potentials. In the sequence, we introduced examples of potentials with polynomial and hyperbolic interactions, classified in two distinct families of models,  which were deformed by a function containing a parameter that controls the smoothness of the deformed potential. It was found that if the starting model has asymmetric topological sectors, then a new sector with modified stability potential can arise on the deformed one, besides those with the same stability potential as the starting model. 

Furthermore, we also analyzed the kink solutions and energy densities of these systems. The several results of the current work consolidate the motivation included in Refs.~\cite{BEN2,BL}, that there may be distinct field theory models that engender the very same stability potential which, in this way, gives rise to the same quantum mechanical potential. As an extra result, we observe that for specific choices of the models, both starting and deformed potentials can define twinlike models \cite{Andrews}, since they develop the same energy densities, stability potentials, and static solutions shifted by a constant. The issue concerning twinlike models will be further examined elsewhere and we are also dealing with other localized structures, in particular, with the possibility to extend the work to the case of planar structures like vortices and skyrmions.

\centerline{*****}

The authors would like to thank CAPES and CNPq for partial financial support.


\end{document}